\documentclass[twocolumn, pre, aps, 10pt]{revtex4-1}
\pdfoutput=1
\usepackage[english]{babel}
\usepackage[utf8x]{inputenc}
\usepackage[caption=false]{subfig}
\usepackage{graphicx}
\usepackage{amsmath,amssymb}
\usepackage{natbib}
\usepackage{algorithm}
\usepackage{algpseudocode}
\usepackage{tabularx}
\usepackage{color}

\newcommand{\x}{\vec{x}}
\DeclareMathOperator*{\argmin}{\arg\!\min}
\DeclareMathOperator*{\opt}{\arg\!opt}

\begin{document}

\title{Prediction of Dynamical Systems by Symbolic Regression}

\author{Markus Quade}
\affiliation{Universit\"at Potsdam, Institut f\"ur Physik und Astronomie, Karl-Liebknecht-Stra{\ss}e 24/25, 14476 Potsdam, Germany}
\affiliation{Ambrosys GmbH, David-Gilly Stra{\ss}e 1, 14469 Potsdam, Germany}

\author{Kamran Shafi}
\affiliation{School of Engineering and Information Technology,
The University of New South Wales,
Canberra ACT 2600,
Australia}

\author{Bernd R.\ Noack}
\affiliation{Laboratoire d'Informatique pour la M\'ecanique et les Sciences de l'Ing\'enieur LIMSI-CNRS,  BP 133,  91403 Orsay cedex, France}
\affiliation{Institut f\"ur Str\"omungsmechanik,
Technische Universit\"at Braunschweig,
Hermann-Blenk-Stra{\ss}e 37,
38108 Braunschweig, Germany
}
\author{Robert K.\ Niven}
\affiliation{School of Engineering and Information Technology,
The University of New South Wales,
Canberra ACT 2600,
Australia}

\author{Markus Abel}
\affiliation{Universit\"at Potsdam, Institut f\"ur Physik und Astronomie, Karl-Liebknecht-Stra{\ss}e 24/25, 14476 Potsdam, Germany}
\affiliation{Ambrosys GmbH, David-Gilly Stra{\ss}e 1, 14469 Potsdam, Germany}

\begin{abstract}
We study the modeling and prediction of dynamical systems based on conventional models derived from measurements. Such algorithms are highly desirable in situations where the underlying dynamics are hard to model from physical principles or simplified models need to be found.
We focus on symbolic regression methods as a part of machine learning. These algorithms are capable of learning an analytically tractable model from data, a highly valuable property. Symbolic regression methods can be considered as generalized regression methods. We investigate two particular algorithms, the so-called fast function extraction which is a generalized linear regression algorithm, and genetic programming which is a very general method. Both are able to combine functions in a certain way such that a good model for the prediction of the temporal evolution of a dynamical system can be identified. We illustrate the algorithms by finding a prediction for the evolution of a harmonic oscillator based on measurements, by detecting an arriving front in an excitable system, and as a real-world application, the prediction of solar power production based on energy production observations at a given site together with the weather forecast.
\end{abstract}

\date{\today}
\maketitle

\section{Introduction}
\label{intro}

The prediction of the behavior of dynamical systems is of fundamental importance in all scientific disciplines. Since ancient times, philosophers and scientists have tried to formulate observational models and infer future states of such systems. Applications include topics as diverse as weather forecasting \cite{9780511817496}, the prediction of the motion of the planets \cite{gauss1809theoria}, or the estimation of quantum evolution \cite{Schroedinger26}.
The common ingredient of such systems - at least in natural sciences - is the existence of an underlying mathematical model which can be applied as the predictor. In recent years, the use of artificial intelligence (AI) or machine learning (ML) methods have complemented the formulation of such mathematical models through the application of advanced data analysis algorithms that allow accurate estimation of observed dynamics by learning automatically from the given observations and building models in terms of their own modelling languages. Artificial Neural Networks (ANNs) are one example of such techniques that are popularly applied to model dynamic phenomena. ANNs are structured as networks of soft weights organized in layers or so-called neurons or hidden units. One problem of ANN type approaches is the difficult-to-interpret black-box nature of the learnt models. Symbolic regression-based approaches, such as Genetic Programming (GP), provide alternative ML methods that are recently gaining increasing popularity. These methods, similar to other ML counterparts, learn models from observed data and act as good predictors of the future states of dynamical systems. Their added advantages over other methods include the interpretable nature of their learnt models and a flexible and weakly-typed \cite{cardelli1985understanding} modelling language that allows them to be applied to a variety of domains and problems.  

Undoubtedly, the methods used most often in ML are neural networks. These involve deep learning, in the sense that several layers are used and interpreted as the organization of patterns, as one imagines the human brain to work.
In the present study, involving deterministic systems, we want to use a certain branch of ML, namely symbolic regression. This technique joins the classical, equation-oriented approach with its computer-scientific upstart. In this publication we do not present any major improvements in the algorithms; rather we demonstrate how one can apply symbolic regression to identify and predict the future state of dynamical systems.

Symbolic regression algorithms work by exploring a function space, which is generally bounded by a preselected set of mathematical operators and operands (variables, constants, etc.), using a population of randomly generated candidate solutions. Each candidate solution encoded as a tree essentially works as a function and is evaluated based on its fitness or in other words its ability to match the observed output. These candidate solutions are evolved using a fitness-weighted selection mechanism and different recombination and variation operators. One common problem in symbolic regression is the bloating effect which is caused by excessive lengthening of individual solutions or filling of the population by large number of solutions with low fitness. In this work we use a multi-objective function evaluation mechanism to avoid this problem by including minimizing the solution length as an explicit objective in the fitness function. 

Symbolic regression subsumes linear regression, generalized linear regression, and generalized additive models into a larger class of methods. Such methods have been used with success to infer equations of dynamical systems directly from data \cite{Voss-Kolodner-Abel-Kurths-99,abel2005additive,Abel-04,Rey2014869,brunton2015discovering}. One problem with deterministic chaotic systems is the sampling of phase space using embedding.
For a high-dimensional system, this leads to prohibitively long sampling times. 
Typical reconstruction methods use delay coordinates and the associated differences, this results in mapping models for the observed systems. Mathematically, differential coordinates are better suited for modelingbut they are not always accessible from data. Both approaches, difference and differential embedding, are discussed in \cite{Ahnert-Abel-07} with numerical methods to obtain suitable differential  variables from data. Modern methods like diffusion maps \cite{Tenenbaum22122000,singer2009detecting} or local linear embedding \cite{Roweis22122000}, including the analysis of stochastic systems, circumvent the curse of dimensionality by working directly on the manifold of the dynamical system.

Apart from prediction and identification of dynamical systems \cite{schmidt2009distilling,rodriguez1998multi}, the symbolic regression approach has been used recently for the control of turbulent flow systems \cite{Gautier-Abel-Duriez-Segond-Abel-15,Brunton-Noack-2015}.
In that application, we demonstrate how to find the symbolic equations in a very general form combined with subsequent automatic simplification and multiobjective optimization. This yields interpretable equations of a complexity that we can select. We use open-source Python packages for the analysis. Symbolic regression is conducted using an elastic net method provided by the fast function extraction package (FFX) for quick tests, and the more general, but usually slower method implemented as a genetic programming algorithm (GP) based on the deap package. Subsequent simplification is obtained using sympy. Of course, any other programming framework with similar functionality will do. 

For a systematic study we examine numerically-generated data from a harmonic oscillator as the simplest system to be predicted, and a more involved system of coupled FitzHugh-Nagumo oscillators, which are known to produce complex behaviour and may serve as a very simple model for neurons. We investigate the capacity of the ML approach to detect an incoming front of activity, and give exact equations for the regression. We compare different sampling and spatio-temporal embedding methods, and discuss the results: it is shown that a space-time embedding has advantages over time-only and space-only embedding.

Our final example concerns a real-world application, the short-term and medium-term forecasting of solar power production. In principle, this could be achieved trivially by a high-resolution weather forecast and knowledge of the transfer of solar energy to solar cells, a very well-understood process \cite{lorenzo1994solar}. However, such a highly resolved weather forecast does not exist, because it is prohibitively expensive: even the largest meteorological computers are still unable to compute the weather on small spatial scales, let alone with a long time horizon at high accuracy. As the dynamical systems community identified a long time ago, this is mainly due to uncertainties in the initial conditions, as demonstrated by the celebrated Lorenz equations \cite{Lorenz-63}. Consequently, we follow a data-based approach and improve upon weather predictions using local energy production data as a time series. We are aware that use of the full set of weather data will improve the reported forecast, but increasing the resolution is not our interest here, rather the proof of concept of the ML method and its applicability to real-world problems.

The rest of this paper is organized as follows. In Sec.~\ref{sec:methods} we discuss the methods and explain our approach. This section is followed by a longer section \ref{sec:results} where results are presented for the above-mentioned example systems. We end the paper with a summary and conclusions, Sec.~\ref{sec:conclusion}.

\section{Methods}
\label{sec:related_work}

In the field of dynamical systems (DS), and in particular nonlinear dynamical systems, reconstruction of the characteristics of an observed system from data has been and is a fundamental scientific topic. 

In this regard, one can distinguish parameter and structure identification. We first discuss the existing literature on parameter identification which is easier in that there is an established mathematical framework to fit coefficients to known curves representing experimental data, which in turn result from known dynamics. This can be conducted for linear or non-linear functions. For deterministic systems, with the advent of modern computers, quantities like fractal dimensions, Lyapunov exponents and entropies can also be computed to make systems comparable in dynamics \cite{Kantz-Schreiber-97,Ott-94}. These analyses further allow the rough characterization of the type and number of orbits of a DS \cite{strogatz2014nonlinear}. On the other hand, embedding techniques have been developed to reconstruct the dynamics of a high-dimensional system from lower-dimensional time series \cite{Whitney36,Takens-81,Sauer-Yorke-Casdagli-91}.  

These techniques have a number of limitations with respect to accuracy and the amount of data needed for making good predictive models. A chaotic system with positive Lyapunov exponents has a prediction horizon which depends heavily on accuracy and precision of the data, since chaos ``destroys'' information. This can be seen very clearly by the shift map example \cite{Ott-94}. However a system on a regular orbit, even marked with complicated equations, might be predicted accurately. For high-dimensional systems, one needs a large amount of data to address the ``curse of dimensionality'' \cite{Kantz-Schreiber-97}. In fact it can be shown that for each dimension, the number of data needed increases on a power-law basis \cite{Kantz-Schreiber-97,Abarbanel-97}.
Eventually, the direct inference of the underlying equations of motion from data can be approached using regression methods, like Kalman filtering, general linear models (GLM), generalized additive models (GAM), or more general schemes, see \cite{gershenfeld1999nature} and references therein. Apart from the equations themselves, partial derivatives often have to be estimated \cite{Ahnert-Abel-07}, which is an additional problem for low-precision data

We also consider structure identification, which as mentioned above is a more complicated task. In the last 10-15 years, powerful new methods from computer science have been applied to this purpose. This includes numerous studies on diffusion maps, local linear embedding, manifold learning, support vector machines, artificial neural networks, and symbolic regression \cite{Roweis22122000,Tenenbaum22122000,bishop2006pattern,schmidt2009distilling}. Here, we focus on symbolic regression. It must be emphasized that most methods are not unique and their success can only be tested based on their predictive power.

\subsection{Symbolic Regression}

One drawback of many computational-oriented methods is the lack of equations that can be analyzed mathematically in the neighborhood of analyzed trajectories. Symbolic regression is a way to produce such equations.
It includes methods that identify the structure or parameters of the searched equation or both of them simultaneously with respect to objective functions $\Gamma_i$.

This means that methods like GLM, or GAM are contained in such a description. A recent implementation of GLMs is Fast Function Extraction (FFX) \cite{mcconaghy2011ffx}, which is explained briefly below. 
Genetic programming, explained in detail below, is another intuitive method and often used for symbolic regression. Here, the algorithm searches the function space through random combinations and mutations of functions, chosen from a basic set of equations.

Symbolic regression is supposed to be form free and thus unbiased towards human perception. However, human knowledge enters in the meta-rules imposed on the model through the basic building blocks and rules on how they can be combined. Thus, the optimal model is always conditioned on the underlying meta-rules.

\subsubsection{Genetic Programming}

\begin{figure}
\includegraphics[width=\linewidth]{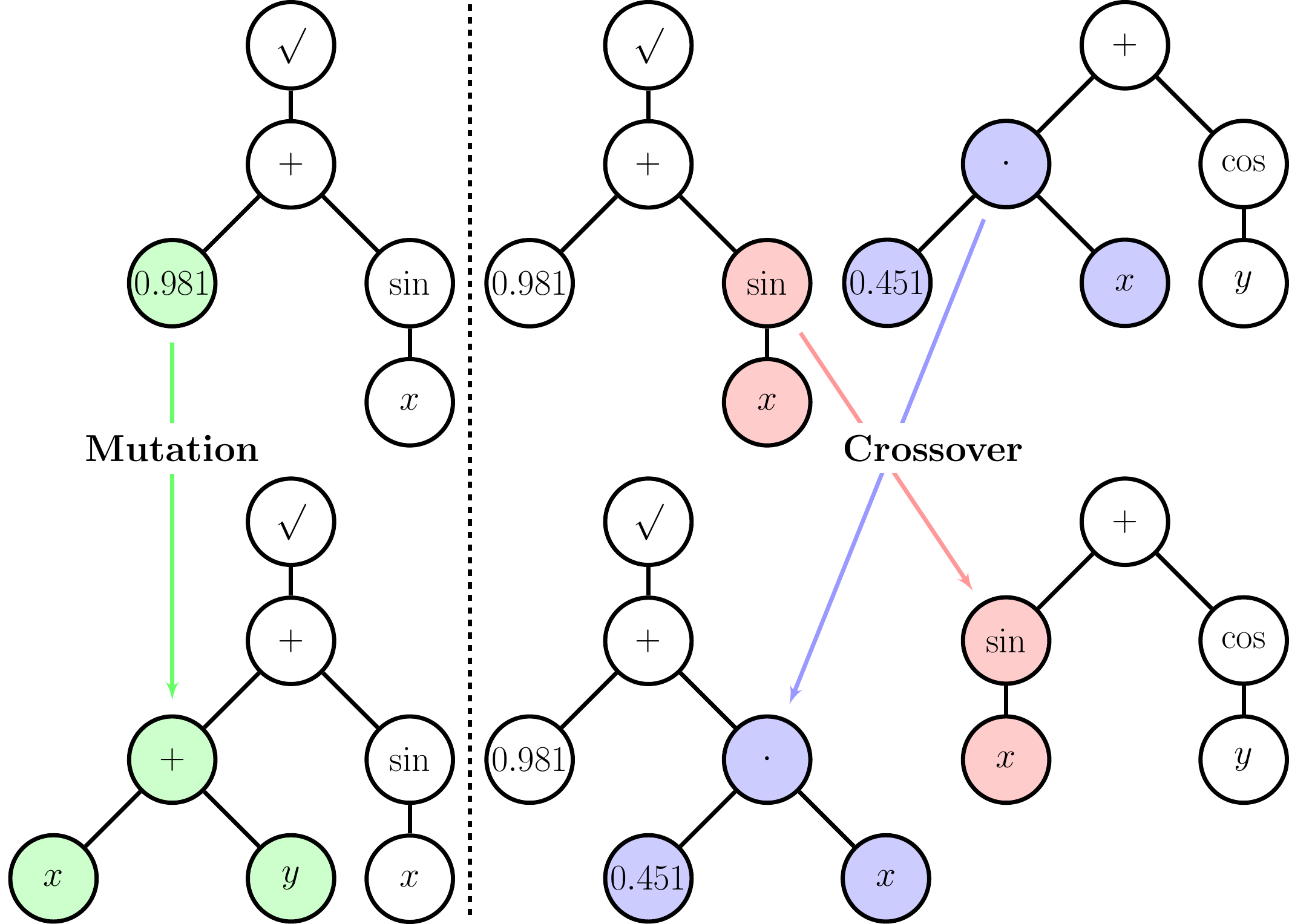}
\caption{Illustration of the genetic programming mutation and crossover. The upper left expression tree describes the function $f(x,y) = \sqrt{0.981 + \sin(x)}$. Mutation is conducted by picking a random subtree, here the single terminal node $0.981$ and replacing it with a new random expression tree. Similarly, the crossover operator (right) takes two expression trees and swaps two random subtrees.}
\label{fig:CrossoverMutation}
\end{figure}

Genetic programming is an evolutionary algorithm to find an optimal algorithm or program. The term ``programming'' in optimization is used synonymously with ``plan'' or algorithm. It was used first by Dantzig, the inventor of linear programming, at a time when computer programs did not exist as we know them today \cite{dantzig1985mathematical}. The algorithm seeks an optimal algorithm, in our case a function, using evolutionary, or ``genetic'' strategies, as explained below. The pioneering work was established by \cite{koza1992genetic}. We can briefly describe it as follows:
in GP we can represent formulae as expression trees, such as that shown in Fig.~\ref{fig:CrossoverMutation}. Non-terminal nodes are filled with elements from a basic function set defined by the meta-rules. Terminal nodes consist of variables or parameters. 
Given the optimization problem 
\begin{equation}
f^* = \opt_{f} \Gamma
\end{equation}
we  seek the optimal solution $f^*$ through optimizing (minimizing or maximizing, or for some cost functionals, finding the supremum or infimum) the fitness (or cost) functional $\Gamma$. To find the optimal solution, GP uses a whole population of candidate solutions in parallel which are evolved iteratively through fitness proportionate selection, recombination and mutation operations. The initial generation is created randomly. Afterwards, the algorithm cycles through the following loop until it reaches its convergence or stopping criteria:
\begin{itemize}
\item \textbf{breed}: Based on the current generation $G_t$, a new set of size $\lambda$ of alternative candidate solutions, the offspring $O_t$, are selected. Several problem-dependent operators are used for this tweaking step, e.g. changing parts of a candidate solution (mutation) or combining two solutions into two new ones (crossover). These tweaking operations may include selection pressure, so that the ``fitter'' solutions are more likely to produce offspring.
\item \textbf{evaluate}: The offspring $O_t$ are evaluated, i.e. their fitness is calculated.
\item \textbf{select}: Based on the fitness value, members of the next generation are selected.
\end{itemize}
This scheme fits the requirements of symbolic regression. 
Mutation is typically conducted by replacing a random subtree by a new tree. Crossover takes two trees and swaps random subtrees between them. This procedure is illustrated in Fig.~\ref{fig:CrossoverMutation}. The fitness function uses a typical error metric, e.g. least squares or normalized root mean squared error.

The random mutations sample the vicinity of their parent solution in function space. As a random mutation could likely lead to less optimal solution, it does not ensure a bias towards optimality. However, this is achieved by the selection, because it ensures that favourable mutations are kept in the set while others are not considered in further iterations.

By design and when based on similar meta-rules, GP includes other algorithms like GLMs or linear programming \cite{bishop2006pattern}.

\begin{algorithm}[H]
\begin{algorithmic}
\Procedure{main}{}
\State $G_0 \leftarrow \text{random}(\lambda)$
\State $\text{evaluate}(G_0)$
\State $t \leftarrow 1$
\Repeat
\State $O_t \leftarrow \text{breed}(G_{t-1}, \lambda)$
\State $\text{evaluate}(O_t)$
\State $G_{t} \leftarrow \text{select}(O_t, G_{t-1}, \mu)$ 
\State $t \leftarrow t + 1$
\Until{$t > T \textbf{ or } G_t = \text{good}()$}
\EndProcedure
\end{algorithmic}
\caption{Top level description of a GP algorithm}\label{alg:mumu}
\end{algorithm}

\subsubsection{FFX and the Elastic Net}
\label{sec:ffx}
Here we briefly summarize the FFX algorithm of McConaghy et al. \cite{mcconaghy2011ffx}. This is a symbolic regression algorithm based on a combined generalized linear model and elastic net approach:
\begin{equation}
f(\x) = a_0 + \sum\limits_{i=1}^{N_B} a_i \phi_i(\x)
\end{equation}
where $\{a_i\}$ are a set of coefficients to be determined, and $\{\phi_i\}$ are an overdetermined set of basis functions described by an heuristic, simplicity-driven set of rules (e.g. highest allowed polynomial exponent, products, non-linear functions, $\ldots$).

In the elastic method, a least squares criterion is used to solve the fitting problem. To avoid overfitting, i.e. high model sensitivity on training data, two regulating terms are added: The $\ell_1$, and $\ell_2$ norms of the coefficient vector. The $\ell_1$ norm favors a sparse model (few coefficients) and simultaneously avoids large coefficients. The $\ell_2$ norm ensures a more stable convergence as it allows for several, possibly correlated variables instead of a single one. The resulting objective function written in its explicit form reads \cite{Zou05regularizationand}:
\begin{equation}
\label{eq:elasticnet}
\vec{a}^* = \argmin_{\vec{a}} ||y - f(\x, \vec{a}) ||_{2} + \lambda \rho ||\vec{a}||_{1} + (1-\rho) \lambda  ||\vec{a}||_{2}
\end{equation}
where $y$ are the data, $\lambda \geq 0$ the regularization weight and $\rho \in [0, 1] $ is the mixing between $\ell_1$ and $\ell_2$ norms.
A benefit of the regularized objective function is that it implicitly gives rise to models with different complexity, i.e. different number of bases $N_B$. 

For large values of $\lambda$, the predicted coefficients will all be zero. Reducing $\lambda$ will result in more complicated combinations of non-zero coefficients. For every point on the $(\lambda, \rho)$-grid, the ``elastic net'', one can obtain a single optimal model using a standard solver like coordinate descent to determine the optimal coefficients $\vec{a}^*$.

A small change in the elastic net parameters leads to a small change in $\vec{a}^*$ such that one can use the already obtained solution of a neighboring grid point to restart coordinate descent with the new parameters. 
 
For the obtained models we can calculate the normalized root mean-squared error and model complexity (number of used basis functions). 
The FFX algorithm is based purely on deterministic calculations. Hence its runtime compared to a similar GP algorithm is significantly shorter. However, the meta-rules are more stringent.

\subsection{Multiobjective Fitness}
\label{sec:multiobjectivity}

As mentioned above, the solution of the regression problem is not unique in general. 
A major factor which motivates symbolic regression is its comprehensible white-box nature opposed to the black-box nature of, for example neural networks. Invoking Ockhams razor (lex parsimoniae), a simple solution is considered superior to a complicated one \cite{blumer1987occam,jefferys1992ockham} as it is more easy to comprehend. In addition, more complicated functions are prone to overfitting.
This means that complexity should be a criterion in the function search, such that more complex functions are considered less optimal. We therefore seek a solution which satisfies two objectives.

Comparing solutions by more than one metric $\Gamma_i$ is not straightforward.
One possible approach is to weight these metrics into one objective $\Gamma$:

\begin{equation}
\Gamma = \sum\limits_i^N w_i \Gamma_i
\end{equation} 
making different candidate solutions easily comparable. The elastic net Eq.~\ref{eq:elasticnet} uses such a composite metric.
However, \textit{a priori} it is assumed that there is a linear trade-off between the individual objectives. This has three major flaws:

\begin{itemize}
\item One needs to determine suitable (problem dependent) $w_i$.
\item One does not account for non-linear trade-offs (e.g. all-or-nothing in one objective).
\item Instead of single optimal solution there may be a set of optimal solutions defining the compromise between conflicting objectives (here error vs complexity).
\end{itemize}

\noindent The optimal set is also called the Pareto-front. This is the set of non-dominated candidate solutions, i.e. candidate solutions that are not worse than any other solution in the population when compared on all objectives. For the FFX algorithm, explained above, one can obtain the (Pareto-) optimal set of candidate solutions by sorting the models. The mapping from parameter space to the Pareto-optimal set is called Pareto-filtering. 

Interestingly, the concept of non-domination already partly solves the sorting problem in higher dimensions as it maps from $\mathcal{R}^N$ to $M$ ordered one-dimensional manifolds: Candidate solutions in the Pareto-front are of rank 0. Similarly, one can find models of rank 1, i.e. all models that are dominated only once (or in other words the nondominated models of all models taken out of the original Pareto-front).

Model 1 $f_1$ can be said to be better than Model 2 $f_2$ if its rank is lower:
\begin{equation*}
f_1 \succ f_2 \Longleftarrow \text{rank}(f_1) < \text{rank}(f_2)
\end{equation*}
To compare models of the same rank, one has to introduce an additional heuristic criterion, for which there are several choices \citep{deb2002fast,zitzler2001spea2,knowles1999pareto}. Usually the criterion promotes uniqueness of a candidate solution to ensure diversity of the population to avoid becoming trapped in a local minimum. As the uniqueness of a solution may depend on its representation and is usually costly to compute, often its projection to fitness space is used. This is conducted to ensure an effective spread of candidate solutions on the Pareto-front.

For example, the non-dominated sorting algorithm II (NSGAII) \citep{deb2002fast} uses a heuristic metric called crowding distance or sparsity to compare two models of the same rank. The scaled Euclidean distance in fitness space to the neighboring models is used to describe the uniqueness of a model. For NSGAII we have: 
\begin{align}
f_1 \succ f_2 \Longleftarrow \begin{cases} &\text{rank}(f_1) < \text{rank}(f_2)  \\ \
\begin{split} &\text{rank}(f_1) = \text{rank} (f_2)  \text{ and }\\ &\hspace{20pt} \text{sparsity}(f_1) > \text{sparsity} (f_2) \end{split}
\end{cases} 
\end{align}
Out of the current generation and their offspring $G_t \cap O_t$ the $\mu$ best, in terms of $\succ$, solutions are chosen for the next generation $G_{t+1}$. This selection method ensures elitism, i.e. the best solutions found so far are carried forward in next generations. Looking at the high-level description in Algorithm~\ref{alg:mumu}, $G_t$ can be seen as an archive which keeps old members as long as they are not dominated by a new solution from the current offspring $O_t$.

The different selection strategies were first studied in the context of genetic algorithms, but more recently they have been successfully applied to symbolic regression \cite{smits2005pareto,vladislavleva2009order}.

\section{Our GP setup}
\label{sec:methods}
For all applications below, our function set is $\{+, *, -, /, \sin, \cos, \exp, \log, \sqrt{}, \; ^2\}$. All discontinuities are defined as zero. Our terminal set consists of the input data $x_i$ as well as symbolic constants $c_i$ which are determined during evaluation.
We set up our multiple objectives as follows: 
the algorithm runs until the error of the most accurate model is below 0.1\%, or for 100 generations. The population size $\mu$ as well as the number of offspring per generation $\lambda$ is set to 500. The depth of individuals of the initial populations varies randomly between 1 and 4. With equal probability we generate the corresponding expression trees where each leaf might have a different depth or each leaf is forced to have the same depth. For mutation we randomly pick a subtree and replace it with a new tree, again using the half and half method, with minimum size 0 and maximum size 2. Crossover is conducted by randomly picking a subtree each and exchanging them. 
Our breeding step is composed of randomly choosing two individuals from the current population, performing crossover on them with probability $p=0.5$ and afterwards always mutating them. Our multiobjective cost functional has the following components 
\begin{equation}
\Gamma_1 = \text{NRMSE}\left(y, \hat{y}\right) = \dfrac{\sqrt{\sum\limits_{i=1}^N \dfrac{(y_i - \hat{y}_i)^2}{N}}}{y_{max} - y_{min}}
\label{eq:NRMSE}
\end{equation}
where NRMSE is the normalized root mean-squared error of the observed data $y$ and its predictor $\hat{y} = f(\vec{x})$, and
\begin{equation}
\Gamma_2 = \text{size}(f)
\label{eq:complexity}
\end{equation}
is simply the total number of nodes in the expression tree $f$. Selection is conducted according to NSGAII. 
In this paper, a model is called accurate if its error metric $\Gamma_1$ is small, where ``small'' depends on the context. For example, numerical data might be modeled accurately if $\Gamma_1 \le 0.05$ and measured data might be modeled accurately if $\Gamma_1 \le 0.20$.  Similarly a model is complicated if its complexity $\Gamma_2$ is relatively large. ``Good'' and its comparatives are to be understood in the sense of $\succ$.

During the generation of the initial population and selection, we force diversity by prohibiting identical solutions. It is very unlikely to randomly create identical solutions. However, offspring may be nearly identical in structure as well as fitness and consequently a crossover between parent and child solution may produce an identical grandchild solution. The probability of such an event grows exponentially with the number of identical solutions in a population and therefore it reduces the diversity of the population in the long-term risking a pre-mature convergence of the algorithm.
Thus, by prohibiting identical solutions, the population will have a transient period until it reaches its maximum capacity. This will also reduce the effective number of offspring per generation. This change reduces the probability of becoming trapped in a local minimum because of a steady state in the evolutionary loop.

Our main emphasis is the treatment of the model parameters $c_i$. In standard implementations, e.g. the already mentioned \cite{smits2005pareto,vladislavleva2009order}, the parameters are mutated randomly, like all other nodes. Here, using modern computational power we are able to use traditional parameter optimization algorithms. Thus, the calculation of $\Gamma_1$ becomes another optimization task given the current model $f_j$:

\begin{equation}
\Gamma_1 = \text{NRMSE}\left(y, f(\vec{x}, \vec{c}^*)\right)
\end{equation} with 

\begin{equation}
\vec{c}^* = \argmin_{\vec{c}}\text{NRMSE}\left(y, f(\vec{x}, \vec{c})\right)
\end{equation} 

\noindent The initial guess for $c_i$ is either inherited or set to one. Thus, we effectively have two combined optimization layers.
Each run is conducted using 10 restarts of the algorithm. The Pareto front is the joined front of the individual runs.  
Finally, we can use algebraic frameworks to simplify the obtained formulae. This is useful, since a formula (phenotype, macrostate) may be represented by many different expression trees (genotypes, microstates). 

\section{Case Studies} 
\label{sec:results}
We present here results for three systems with increasing difficulty: first, we demonstrate the principles using a very simple system, the harmonic oscillator; second, we infer a predictive model for a set of coupled oscillators; and finally we show how we can predict a very applied system, namely the power production from a solar panel. For the first two examples we use numerically produced data, where we have full control over the system, while for the demonstration of applicability we use data from a small solar power station \cite{unisolar}.

\subsection{Harmonic Oscillator}
In this subsection we describe the first test of our methodology:
an oscillator should be identified correctly and a accurate prediction must be possible. 
Consequently, we investigate the identification of a prediction model, not necessarily using a differential formalism. This might be interpreted as finding an approximation to the solution of the underlying equation by data analysis. A deep investigation of the validity of the solution for certain classes of systems is rather mathematical and is beyond the scope of this investigation.

Our system reads
\begin{eqnarray}
 \dot{x} &=& y \\
 \dot{y} &=& -\omega^2 x 
\end{eqnarray}
where $x$ and $y$ are the state variables and $\omega$ is a constant. We use the particular analytical solution $x(t) = x_0 \sin(\omega t)$, $y(t) = x_0 \omega \cos(\omega t)$. The prediction target is $x(t+\tau)$, where $\tau$ is a time increment. 

Since the analytical solution is a linear combination of the feature inputs, just $N=2$ data points are needed to train the model. This holds for infinite accuracy of the data and serves as a trivial test for the method. In general, a learning algorithm is ``trained'' on some data and the validity of the result is tested on another set, that is as independent as possible. That way, overfitting is avoided. For the same reason one needs to define a stop criterion for the algorithm, e.g. the data accuracy is $10^{-5}$, it is useless and even counterproductive to run an algorithm until a root mean square error of $10^{-10}$ (the cost function used here) is achieved. For the example under consideration, we stop the training once the training error is smaller than 1 

\begin{figure}
 \includegraphics[width=0.9\linewidth]{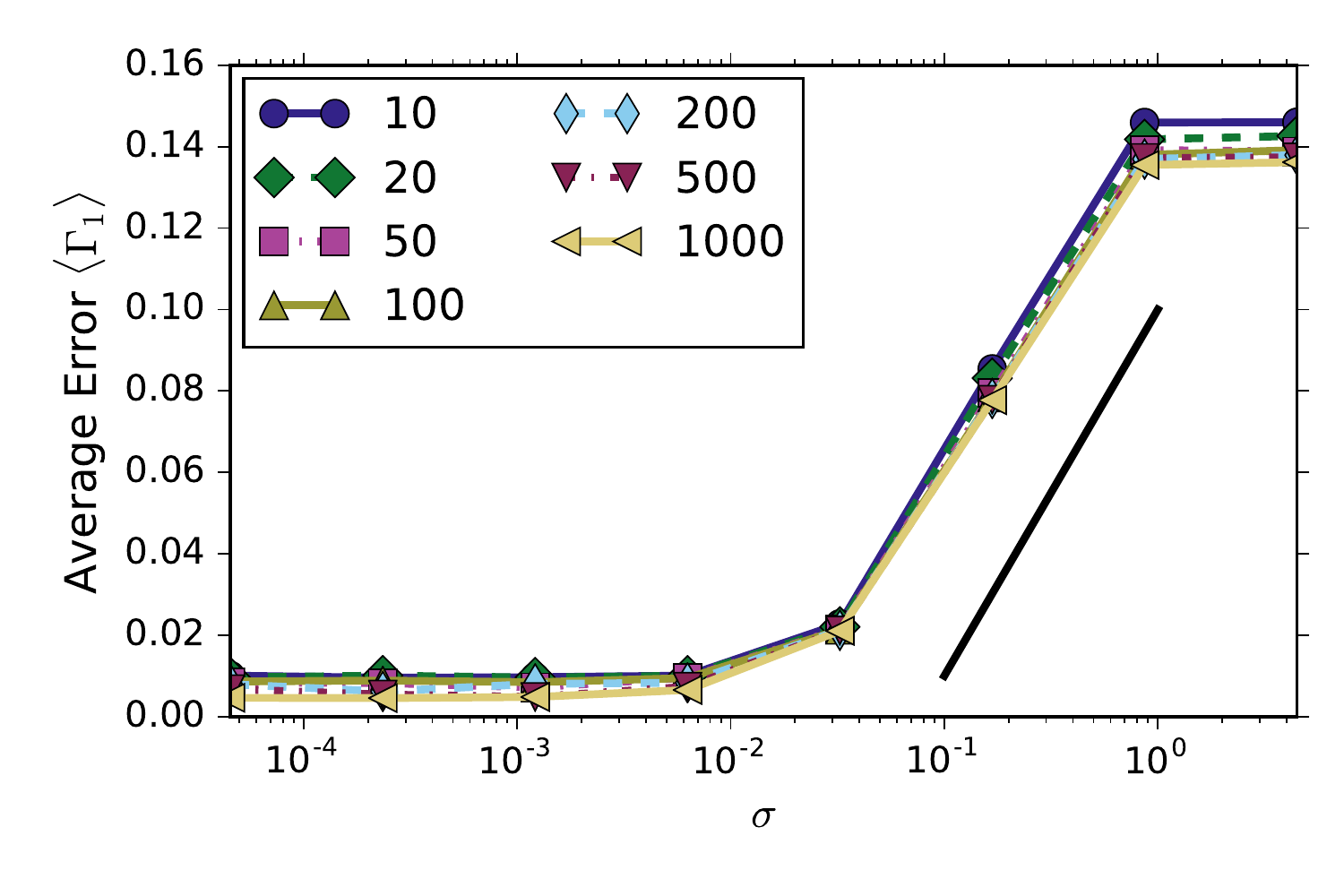}
 \caption{Harmonic oscillator study: NRMSE \eqref{eq:NRMSE} versus noise level $\sigma$ for different training set lengths $N$ and fixed $\tau = 10$. Sufficiently small noise does not worsen the predictability, i.e. the prediction algorithm stops at the target training NRMSE of 1\%. After 0.3 the error does not increase further, since the noise covers the signal completely.}
 \centering
 \label{fig:RmseVsSigma}
\end{figure}

\begin{figure}
 \includegraphics[width=0.9\linewidth]{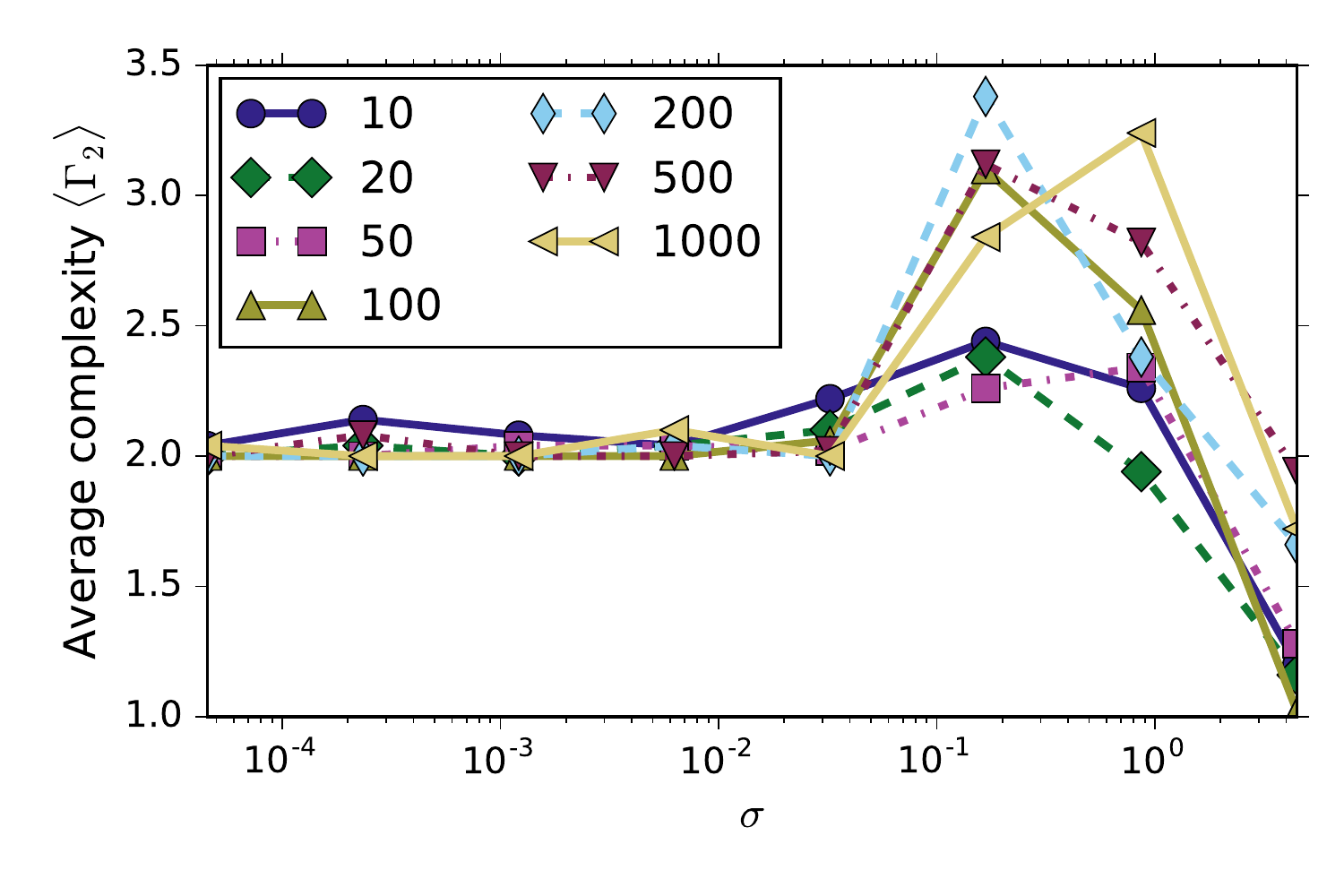}
 \caption{Harmonic oscillator study: Average complexity of the chosen model vs. noise amplitude $\sigma$. The form-free structure allows for overfitting. For small noise, the true solution is found with complexity 2, for higher noise levels, the algorithm starts to fit the noise and more terms are added, reflected by a higher complexity.}
 \centering
 \label{fig:CompVsSigma}
\end{figure}

Typically, a realistic scenario should include the effect of noise, e.g. in the form of measurement uncertainties. 
We consequently add ``measurement'' Gaussian noise with mean zero and variance proportional to the signal amplitude: $\xi_1 \sim \mathcal{N} (0, (\sigma x_0)^2)$, $\xi_2 \sim \mathcal{N} (0, (\sigma x_0 \omega)^2)$, hence $\tilde{x} = x + \xi_1, \tilde{y} = y + \xi_2$.
The training and testing data sets were created as follows: the data are generated between $[0, t_{max}]$. Out of the first half, we chose $N$ values at random for training. For testing purposes we use the second half.
We study the parameter space $(N, \tau, \sigma)$ and average the testing errors over 10 realizations for each parameter set. In Fig.~\ref{fig:RmseVsSigma} we display the normalized root mean squared error of the prediction using FFX (measured against the noisy data) as a function of the noise amplitude. Given $x(t)$ and $y(t)$ the analytical solution for the non-noisy system is just a linear combination, i.e. $x(t+\tau) = \cos(\omega \tau) x(t) + \frac{\sin(\omega \tau)}{\omega} y(t)$, and has a complexity of two. During training we aim for a NRMSE of 1\%. Thus, we find the analytical solution in the limit of small noise amplitude $\sigma$, see Fig.~\ref{fig:RmseVsSigma} and Fig~\ref{fig:CompVsSigma}. 
Strong noise covers the signal and thus the error saturates.

\begin{figure}
 \includegraphics[width=0.9\linewidth]{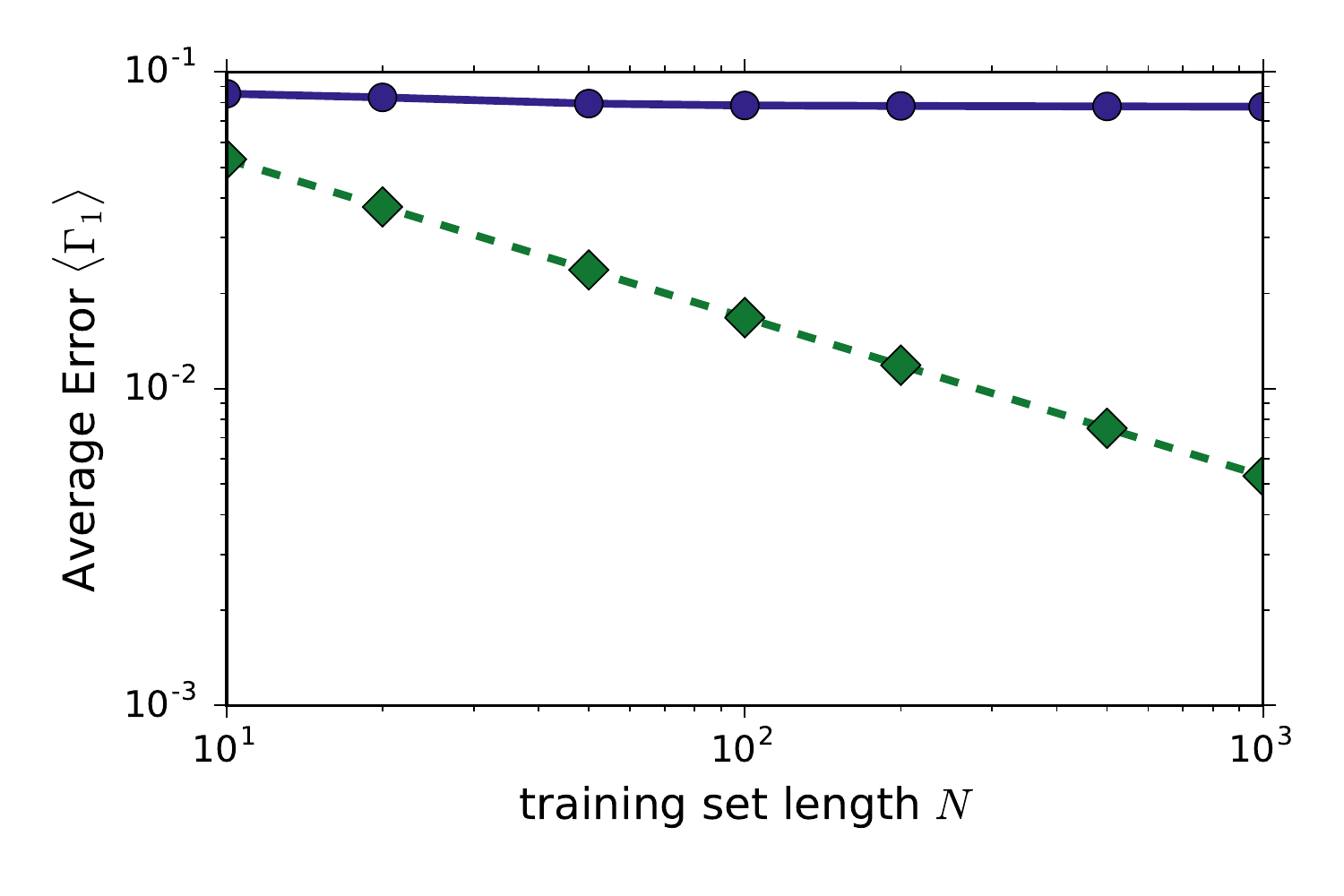}
 \caption{Harmonic oscillator study: In solid blue: normalized root mean squared error vs training set length $N$ for $\sigma=0.17$. Dashed green: $e^{-2}/\sqrt{N}$.
 The error decreases slightly with $N$, but the scaling is much less rapid than $1/\sqrt{N}$.}
 \centering
 \label{fig:RmseVsLength}
\end{figure}

The length of the analyzed data is another important parameter: typically one expects convergence of the error $ \sim \frac{1}{\sqrt{N}}$ for more data. A ``vertical'' cut through the data in Fig.~\ref{fig:RmseVsSigma} is shown in Fig.~\ref{fig:RmseVsLength}. The training set length $N$ has a much lower impact than the classical scaling suggests. Crucial for this scaling is the form free structure as well as the heuristic which is used to select the final model.
For demonstration purposes, we chose the most accurate model on the testing set, which is of course vulnerable to overfitting.
The average complexity, calculated by Eq.~\eqref{eq:complexity} of the final model as a function of the noise amplitude, is shown in Fig~\ref{fig:CompVsSigma}. As evident we can recover the three regimes of Fig.~\ref{fig:RmseVsSigma}. For small noise, the analytical and numerical solution agree. In the intermediate regime we find on average more complex models (in comparison to the analytical solution). Very strong noise hides the signal and a good prediction is impossible. The optimal solution tends to be single constant , i.e. for high $\sigma$ the complexity tends to smaller values as seen in Fig~\ref{fig:CompVsSigma}.
The prediction error has two components: 1) given a structure, noisy data will lead to uncertain parameters and 2) due to the form-free nature of symbolic regression, noisy data will also lead to an uncertain structure, increasing the uncertainty in the parameters.
Thus, final model selection has to be performed carefully, especially when dealing with noisy data. A detailed study is presented for the example of coupled oscillators.

\subsection{Coupled Oscillators}

The harmonic oscillator is an easy case to treat with our methods. 
Now, we extend the analysis to add a spatial dimension. We study a model of FitzHugh-Nagumo oscillators \cite{nagumo1962active} on a ring. The oscillators are coupled and generate traveling pulse solutions.
The model was originally derived as a simplification of the Hodgkin-Huxley model to describe spikes in axons
\cite{hodgkin1952quantitative}, and serves nowadays as a paradigm for excitable dynamics. Here, its spiky behavior is used as an abstraction of a front, observed in real world applications like the human brain, modeled by connected neurons, or a wind power plant network where fronts of different pressure pass through the locations of the wind power plants. The aim is to show that temporal and/or spatial information on the state of some network sites enables an increase in predictability of a chosen site or eventually (if there are waves in the network) to the front detection. The model for the $i$th oscillator is:
\begin{eqnarray}
\dot{v_i} &=& v_i - \dfrac{v_i^3}{3} - w_i + I_i + D\sum\limits_{i,j} A_{ij} (v_j - v_i)  \\
\dot{w_i} &=& \varepsilon (v_i + a - b w_i)\;.
\end{eqnarray}
where $v_i$ and $w_i$, $i,j={1,\dots,N}$, denote the fast and slower state variables, $I_i$ is an external driving force, $D$ is the coupling strength parameter, and $A_{ij} \in \{0,1\}$ describes the coupling structure between nodes $i$ and $j$. The constant parameters $\varepsilon$, $a$ and $b$ determine the dynamics of the system as $\varepsilon^{-1}$ is the time scale of the slower ``recovery variable'', and $a$, and $b$ set the position of the fixed point(s). For $A_{ij}$ we choose diffusive coupling on a ring, i.e. periodic boundary conditions. With the external current $I_i$ we can locally pump energy into the system to create two pulses which will travel with the same speed but in opposite directions, annihilating when they meet.

Using different spatio-temporal sampling strategies, the aim is to detect and predict the arrival of a spike train at a location far enough away from the excitation center (i.e. farther than the wave train diameter). We mark this special location with the index zero.

Note that we do not aim to find a model for a spatio-temporal differential equation, since this would involve the estimation of spatial derivatives, which in turn require a fine sampling. This is definitely not the scope here. Rather we focus on the more application-relevant question to make a prediction based on an equation.

The construction of the data set was similar to the single oscillator case: sensors were restricted to the $v_i$ variables. We can record the time series of $v_0$ and use time delayed features for the prediction. Another option is to use information from non-local sensors.

We prepare and integrate the system as follows: we consider a ring of $N=200$ oscillators. The constants are chosen as $a = 0.7$, $b = 0.8$, $\tau = 12.5$ and $D = 1$. The system is initialized with $v_i(0) = 0$ and $w_i(0) = -1.5$. With the characteristic function $\chi_T(x) = 1 \text{ if } x \in T \text{ else } 0$ we can write the space and time dependent perturbation as $I_i(t) = 5  \chi_{t - \lfloor t \rfloor \le 0.4}(t) \chi_{t \le 40}(t)  \chi_{i \in \{-50, -49\}}(i)$. This periodic perturbation leads to a pair of traveling waves. The data were sampled at times $t_n = n \Delta t$ with $\Delta t = 0.1$. The system has multiple time scales: 
two are associated with the on-site FitzHugh-Nagumo oscillator ($\tau_{fast} = 1$, $\tau_{slow} = \frac{1}{\varepsilon}$), while two more are due to diffusive coupling ($\tau_{Diff} = D$) and perturbation ($\tau_{Pert}$ behaves as $I_i(t)$ described above).
\begin{figure}
\includegraphics[width=0.9\linewidth]{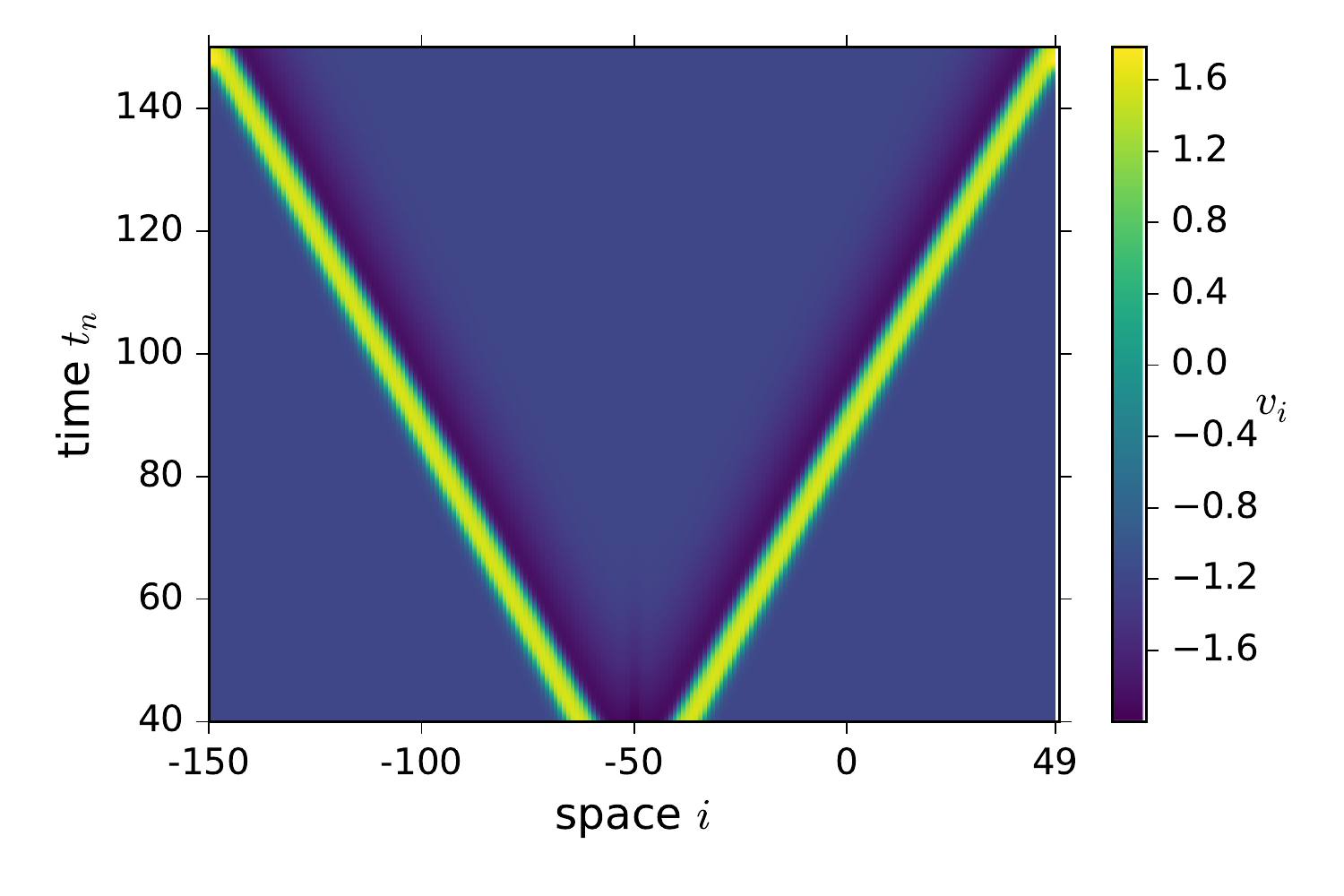}
\caption{Space-time plot of the pulse evolution. $v_i$ is color coded. The front velocity is $v_f = 1.28$. Pulse width (full width half maximum) $\tau_P = 8.4$}
\label{fig:spacetime}
\end{figure}
The temporal width of the pulse traveling through a particular site, $\tau_P = 8.4$, corresponds to the full width half maximum of the pulse. In Fig.~\ref{fig:spacetime} we show the evolution of the oscillator network. The state of $v_i$ is color-coded. The horizontal width of the yellow stripe corresponds to the spatial pulse width $\xi\simeq 10.75$. The speed of the spike or front is consequently $v_{front}\sim \xi/\tau_P = 1.28$. An animation of this can be found in the supplemental material.
The training data, denoted as well feature set, were recorded in three different ways:

\begin{itemize}
\item \textbf{site-only}: Only $v_0$ is recorded, and time-delayed features $v_{0, \Delta n} = v_{0}(t = (n - \Delta n)\Delta t)$ are also included with $\Delta n \Delta t = -1, -2, -3, -4$.
\item \textbf{spatially extended}: We record $v_0$ and additionally $v_{i}$ with $i = -2, -4,\ldots,-10, -20$ (upstream direction).
\item \textbf{mixed}: This combines the two approaches above. For each site we also include the time delayed features.
\end{itemize}
To avoid introducing additional symbols we use state variables with double subscripts for discrete times, where the second index refers to time, and one subscript for continuous time. The respective useage is evident from the context.
We choose to predict the state at time $t=2$ given the data described above. In other words, the prediction target is $v_0 (t_n+\tau)$ with $\tau = 20 \simeq 2.5 \tau_P$, corresponding to the requirement to be far engouh from the excitation point.
Of course, this implies a distance of $\Delta x \sim  2.5 \xi$. The testing and training sets were selected by using every second point of the recorded time series. 

\subsubsection{FFX Results}

\begin{figure}
 \includegraphics[width=0.9\linewidth]{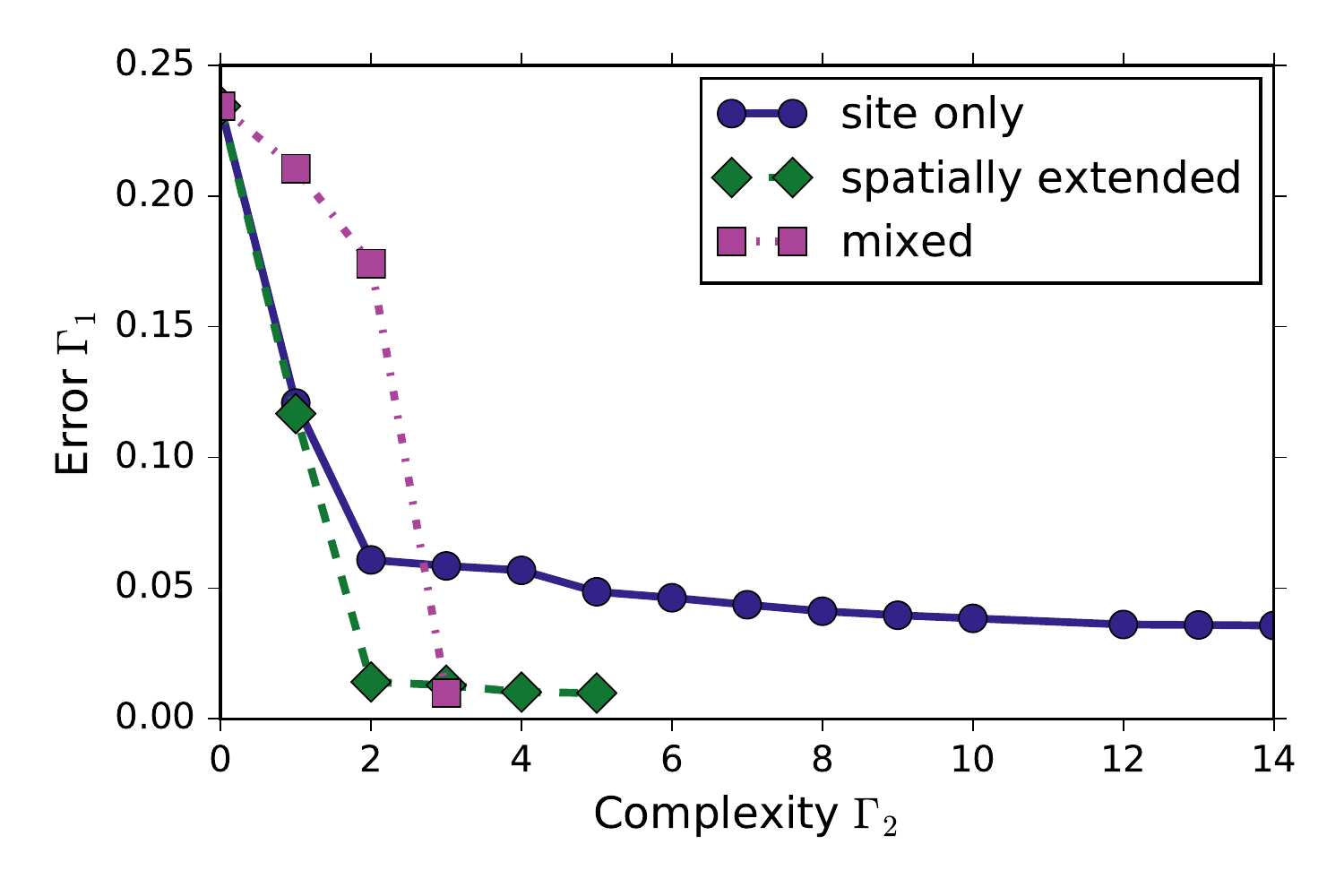}
 \caption{Coupled spiking oscillators, method FFX: Pareto fronts for the different spatio-temporal samplings of the network data. For this plot we use $\rho = 0.95$. This leads to the non-convex shape of the front based on the most information. The models are re-evaluated on the testing set.}
 \centering
 \label{fig:ffxpareto}
\end{figure}

We first discuss the results obtained by FFX (Sec.~\ref{sec:ffx}).
In Fig.~\ref{fig:ffxpareto} we display the Pareto fronts using the three different approaches for the training set. All curves have one point in common which represents the best fitting constant (complexity 0). As one would expect, the site only data do not contain enough information to detect a front. Thus, even high complexity models cannot reach an error below 4\% and the required error of 1\% is never met. In the two other datasets the algorithm has the possibility to find a combination of spatial amd temporal inputs to account for the front velocity. Note that the shape of the front strongly depends on the internal $\rho$ parameter of the elastic net Eq.~\ref{eq:elasticnet}. More information should not lead to a decrease in predictability. Thus, the Pareto front of a data set richer in features dominate the corresponding Pareto front of a data set with less features. Counter-intuitively, using $\rho = 0.95$ \footnote{The default value of the package which works well in most scenarios.} the front for the mixed dataset becomes non-convex as some good fitting models are hidden by the regularizer. Thus, we can use $\rho$ to influence the shape of the front. Despite that, the most accurate model of the mixed data set is still the most accurate model overall.

In the following we discuss the results for the best models for each feature set.

\begin{figure}
\subfloat[]{\includegraphics[width=0.9\linewidth]{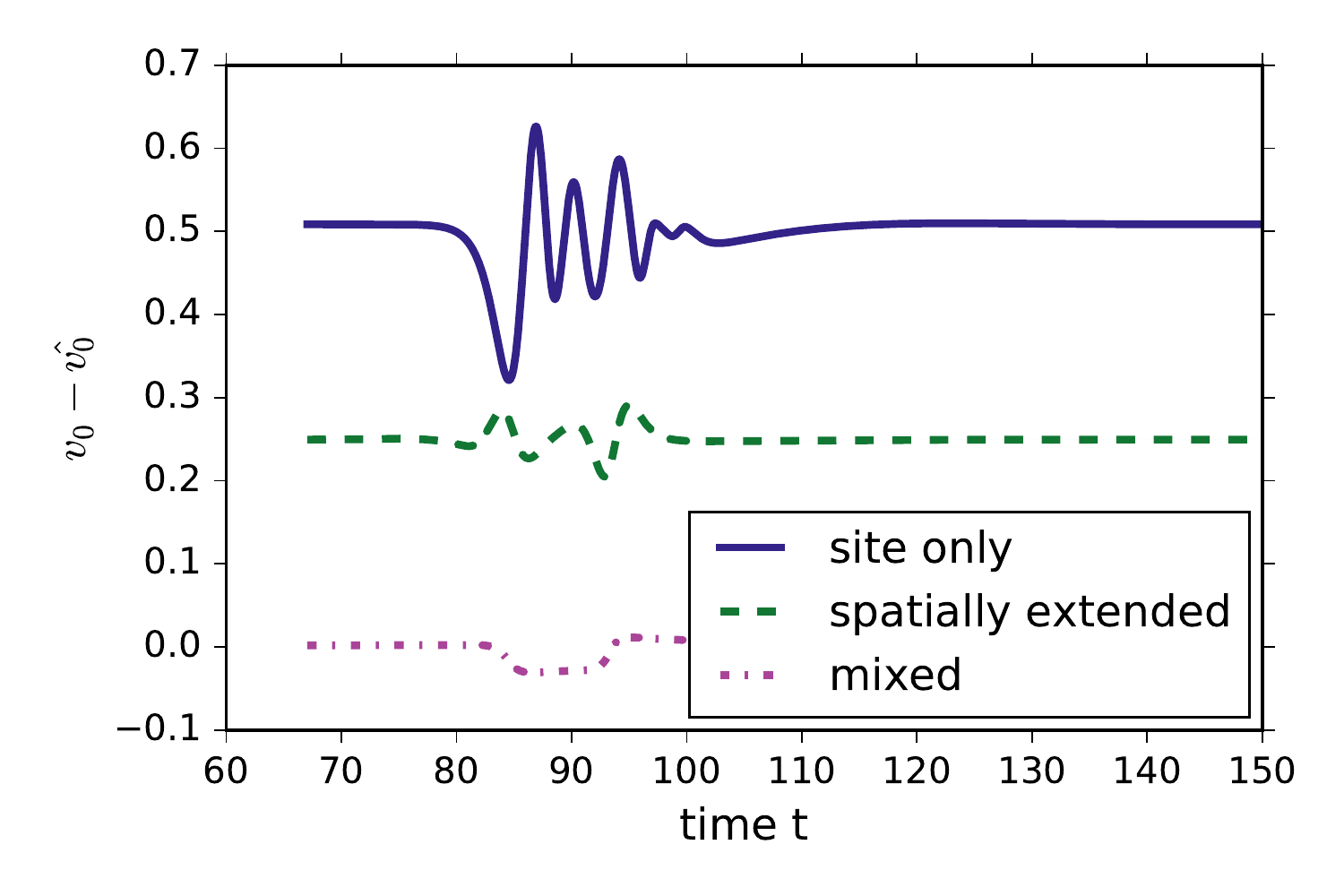}\label{fig:ffx_diffa}}
\\
\subfloat[]{\includegraphics[width=0.9\linewidth]{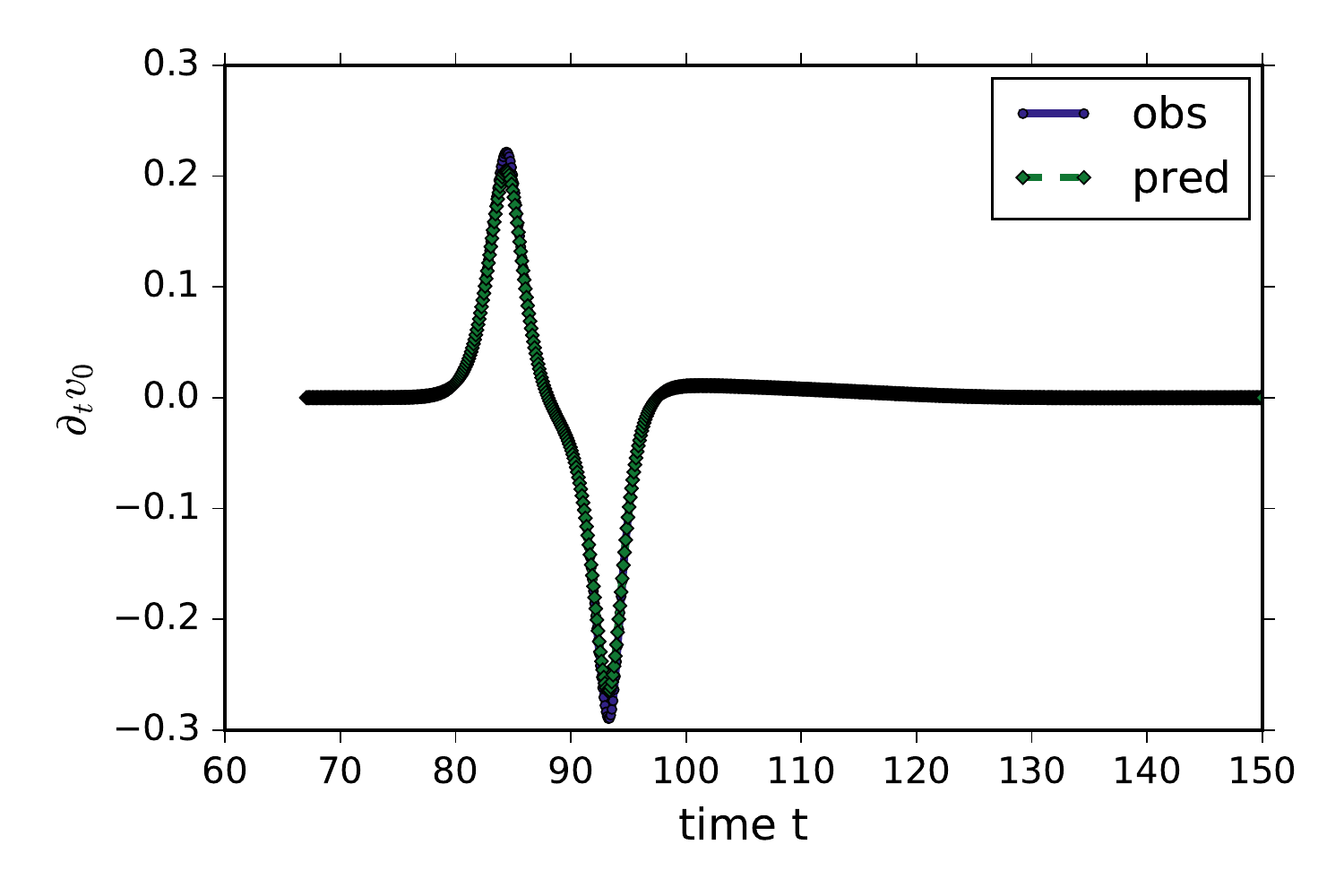}\label{fig:ffx_diffb}}
\caption{Coupled spiking oscillators, method FFX. For each feature set, the most accurate model is used as the predictor $\hat{v}_0$. In (a) we show the difference $\delta v_0 = v_0 - \hat{v}_0$. The upper two curves are shifted by 0.25 and 0.5 respectively. In (b) we compare the time derivative (approximated by the finite difference quotient) of the most accurate model overall and the real data. For details see text.}
\end{figure}

If we take the perspective of an observer sitting at $i=0$, we see the spike passing: first the state is zero, then a slow increase is observed followed by a rapid increase and decrease around the spike maximum. Eventually the state returns slowly to zero. Statistically, the algorithm is trained by long quiet times and a short, complicated spike form which is hard to model by a reduced set of state variables. This is illustrated in Fig.~\ref{fig:ffx_diffa} where for any feature set the biggest differences occur in the spike region. Apparently, the model with site-only variables shows worse results than the spatial one, and the spatio-temporal set models best the passing spike. We note that in a direct confrontation, the true and modeled signal would hard to be distinguished. In Fig.~\ref{fig:ffx_diffb} we confront the time derivative for the model from mixed variables. The true and modeled spike are indistinguishable by eye.

\begin{table*}
\begin{tabularx}{\linewidth}{|l|X|}
\hline 
 temporal site-only & $-0.0273 + 3.34v_{0,0} - 2.41v_{0,0}  v_{0,-10} - 2.09v_{0,-40}  v_{0,-10} + 1.64v_{0,-20}^2 - 1.53v_{0,-20} - 1.16v_{0,-10} + 0.991v_{0,-30}^2 + 0.684v_{0,0}^2 + 0.463v_{0,-30} + 0.433v_{0,-20}  v_{0,0} + 0.373v_{0,-20}  v_{0,-10} - 0.359v_{0,-40}^2 + 0.216v_{0,-40} + 0.00286v_{0,-10}^2$\\ 
 \hline  
spatially extended & $-0.00247 + 0.897v_{-2,0} + 0.178v_{0,0} - 0.0650v_{-4,0} + 0.00280v_{-10,0} - 0.00210v_{-8,0}$\\ 
 \hline 
temporal spatial &$0.00894 + 0.442v_{-4,-30} + 0.346v_{-2,-10} + 0.175v_{-2,0}$\\ 
 \hline
\end{tabularx} 
\centering
\caption{Coupled spiking oscillators, method FFX. Formulae of the most accurate models. The spatio-temporal embedding reproduces the data very well, i.e. an early detection is possible.}
\label{table:ffxeq}
\end{table*}

The formulae of the most accurate models are shown in Table~\ref{table:ffxeq}.
For site-only features, quadratic combinations of points at different times occur. This reflects the approximation of the incoming front by a quadratic term. If, however only spatial points are used, the dynamics far away are used to predict the incoming front. If the small terms are neglected, the model consists of the signal at the target site itself, and the previous site (-2) which carries the largest weight. Physically, it means that despite being far away the front is already felt at 2 sites away. Since the front is stationary in a co-moving frame, spatio-temporal embedding is best, namely sampling the spike train in space and moving in time with the train velocity. Then we have a simple and compact linear dependence as seen in the last row of Table~\ref{table:ffxeq}. Let us inspect the possible physics in the model approximating the constants $a_0, a_1, a_2, a_3$ roughly as 
0, 0.45, 0.35, 0.175 such that $a_2=2a_3$ . We first notice that $\tau_p=8.4\simeq 10$. The last terms can then be recombined to $a_3 v_{-2,-10} + a_3 v_{-2,-10}+v_{-2,0}$ as a mean value of the state with time distance of approximately one typical time scale. The state at $-30$ is at the backside of the front and together the most important information, namely the increase and decrease of the incoming signal is selected by the model. 
Alternatively, since $v(0, t) = v(- v_f \tau_P, t - \tau_P)$ the best model in Table~\ref{table:ffxeq} can be interpreted as the weighted average of the closest combination $(\Delta i, \Delta t)$ to represent the front velocity ($\frac{\Delta i}{\Delta t} = \frac{4}{3}\approx v_f$).
This demonstrates how powerful the algorithm works in selecting important features.

\subsubsection{GP Results}

\begin{figure}
 \includegraphics[width=0.9\linewidth]{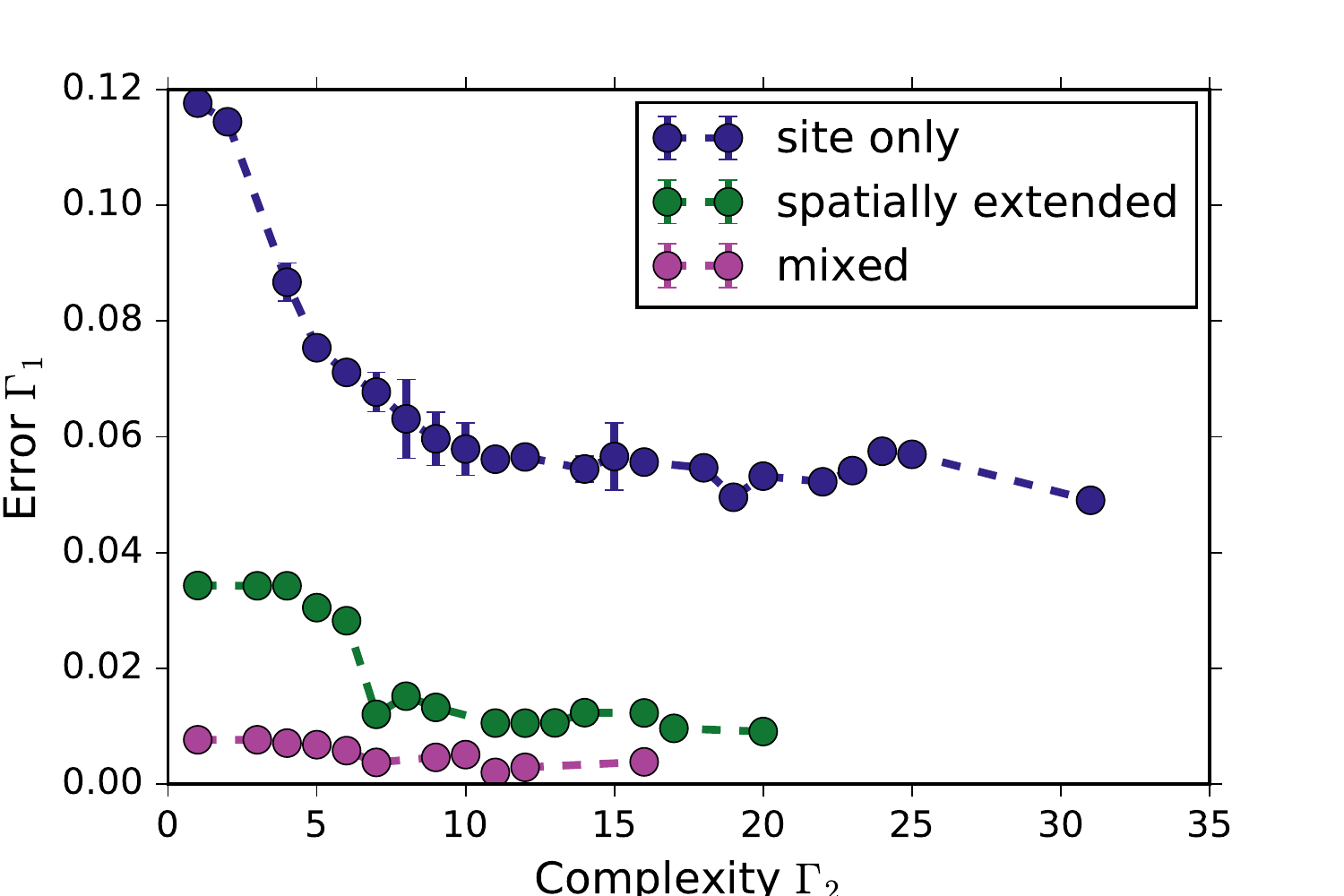}
 \caption{Coupled spiking oscillators, method GP. Averaged Pareto fronts, for each spatio-temporal sampling option, 10 runs are conducted and the resulting complexities and errors are averaged. Errorbars represent the standard deviation. For the spatially extended and mixed data sets the errors are smaller than the circle size. The models are re-evaluated on the testing set.}
 \centering
\label{fig:gppareto}
\end{figure}

We again examine the Pareto-optimal models illustrated in Fig.~\ref{fig:gppareto}. For each feature set we obtain a non-convex Pareto front. The shape and the values of the fronts are broadly similar to the results obtained by FFX. Because GP is an evolutionary method and relies on random breeding rules, we display averaged results: we initialize the algorithm with different seeds of the random number generator, calculate the Pareto fronts and average the errors for the non dominated models of the same complexity. Note that not all complexities occur on each particular front. This way, we obtain a generic Pareto front and avoid atypical models which may occur by chance. The specific model given below in the tables is not averaged, but the best result for one specific seed (42). The errors of the models reachable by the different sets are again decreasing from site only over spatially extended to mixed. However, the mixed model reaches almost zero error which is quite remarkable!

\begin{figure}
\subfloat[]{\includegraphics[width=0.9\linewidth]{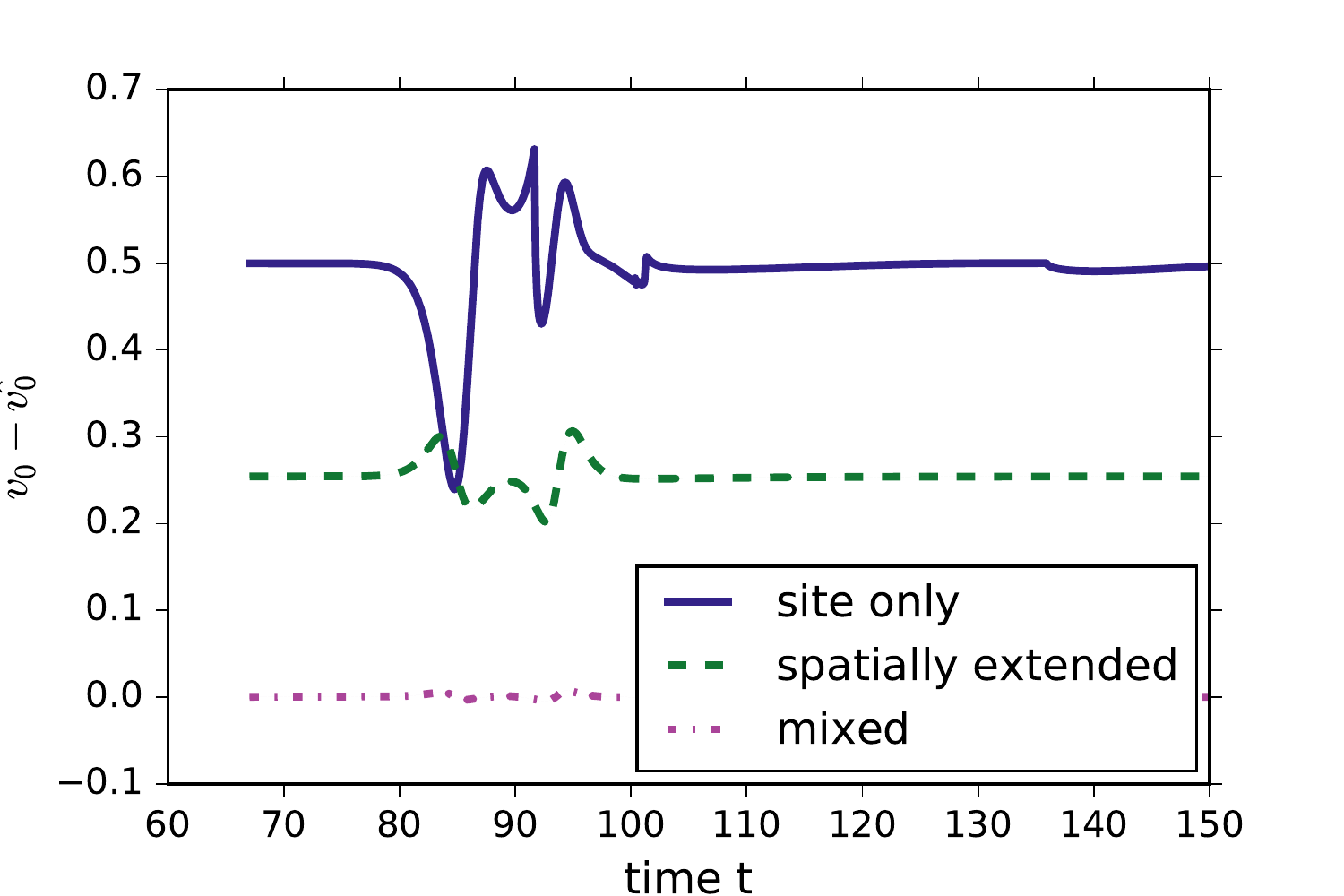}}
\\
\subfloat[]{\includegraphics[width=0.9\linewidth]{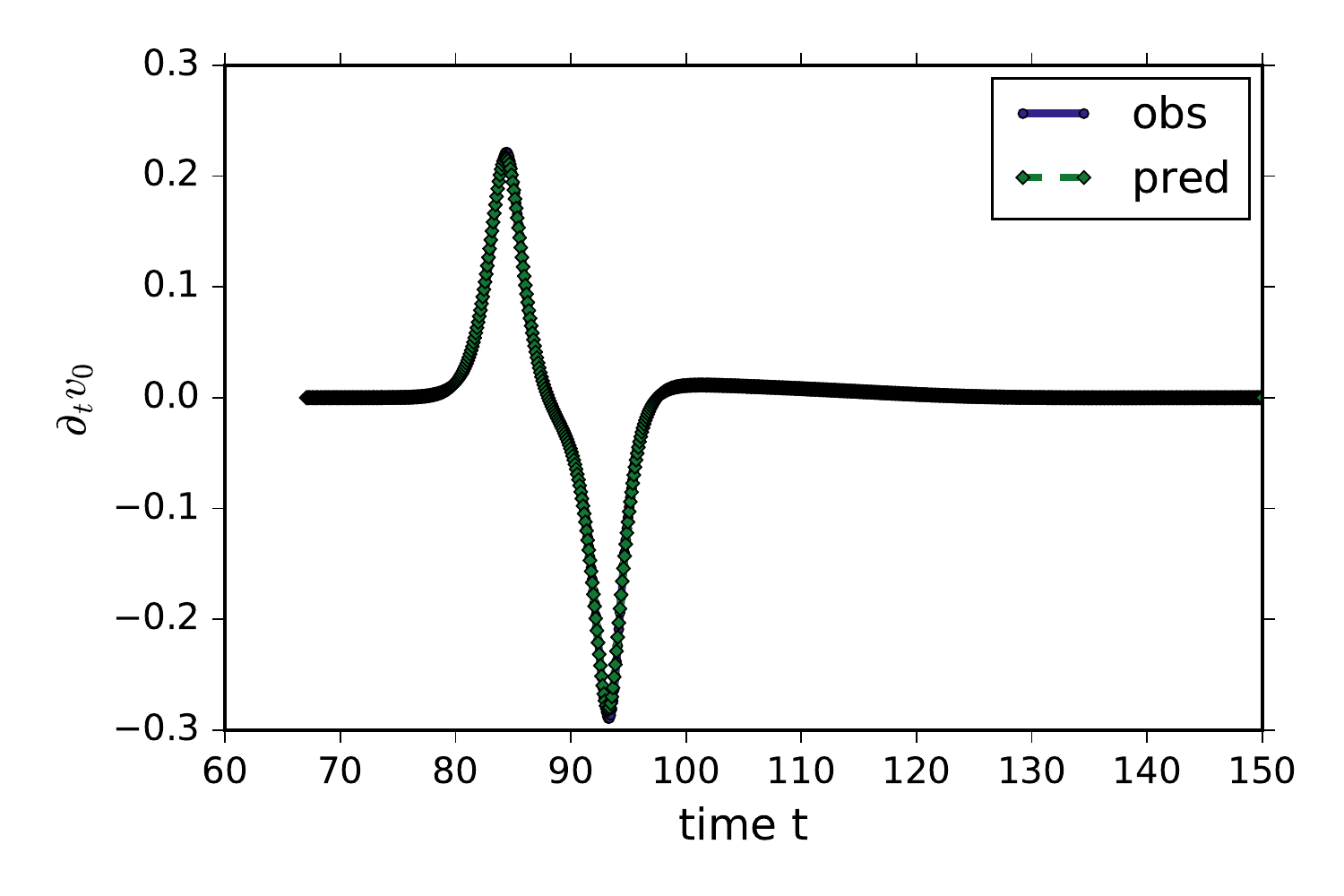}}
\caption{Coupled spiking oscillators, method GP. For each feature set, the most accurate model is used as the predictor $\hat{v}_0$. In (a) we show the difference $\delta v_0 = v_0 - \hat{v}_0$. The upper two curves are shifted by 0.25 and 0.5 respectively. In (b) we compare the time derivative (approximated by the finite difference quotient) of the most accurate model overall and the real data. Prediction and true data cannot be distinguished by eye. }
\centering
\label{fig:gp_diff}
\end{figure}

The difference plots for the method are given in Fig.~\ref{fig:gp_diff}. While the site only set is not able to give a convincing model for an incoming front, the spatially extended set gives a reasonable model with little error. The mixed model is very good with perfect coincidence of model and true dynamics. This model cannot be distinguished by eye from the observed signal.

The models provided by the GP algorithm with seed 42 are given in Table~\ref{table:res_gp}. Due to the very general character of GP these can be overwhelming at first glance. However, we can simplify them down by using computer algebra systems like sympy or mathematica (here we use sympy).

\begin{table*}
\begin{tabularx}{\linewidth}{|l|X|}
\hline 
temporal site-only & $v_{0,0}^2/(v_{0,-10} + \sqrt(-v_{0,-10}(v_{0,0} - v_{0,-30})(v_{0,-30}/\sin(v_{0,-10} + v_{0,-20}) + \exp(v_{0,-30}) - \sqrt(\sin(v_{0,-30})) + \cos(\sqrt(v_{0,-30})v_{0,-40}))))$ \\ 
 \hline 
spatially extended & $0.208v_{0,0} + 0.792v_{-2,0} + 0.0274362547430272\exp(-v_{-4,0})\sin(v_{-2,0})$ \\ 
 \hline 
temporal spatial & $0.878v_{-4,-30} + 0.124496v_{-4,-40}$ \\ 
 \hline 
\end{tabularx} 
\centering
\caption{Coupled spiking oscillators, method GP. Formulas of the most accurate models for seed 42.}
\label{table:res_gp}
\end{table*}

The interpretation of the GP results requires a bit more thinking. In essence, they follow a logic similar to the FFX results. The site-only model is complicated, and instead of a square operator a trigonometric function is used to mimic the incoming pulse. Since the data do not include directly all information needed, the algorithm tries to fit unphysical functions. This is clearly a non-deterministic and overfitting result, mirrored by the high complexity of the functions involved. For spatially extended models, we obtain a linear and sinusoidal components, and the model uses only three features, namely the on-site values and the ones at two and four units left on our site under consideration. Remarkably, a sinusoidal behavior detected with an exponential decrease, which is our intuition.
Eventually, the spatio-temporal embedding yields a very simple model which approximates the front velocity $v_f$ to be between $\frac{4}{3}$ and $1$.
The accuracy of this model is very high.

Summarizing, when given enough input information, both methods find a linear model for the predictor $\hat{v}_0(t+\tau)$ by finding the most suitable combination of temporal and spatial shift to mimic the constant front velocity.
If this information is not available in the input data, nonlinear functions are used.

\subsection{Solar Power Data}

In this section, we describe the results obtained for one-day-ahead forecasting of solar power production. The input data used for training are taken from the unisolar solar panel installation at Potsdam University with about 30 kW installed. Details are found at \cite{unisolar}.
We join the solar power data with meteorological forecast data from the freely available European Centre for Medium-Range Weather Forecasts (ECMWF) portal \cite{ecmwf} as well as the actual observed weather data. These public data are of limited quality and serve for our proof of concept with real data and all their deficiencies.

The solar panel data $P(t)$ were recorded every five minutes, at geoposition 52.41 latitude, 12.98 longitude. The information about the weather can be split into two categories: weather observations of a station near the power source $W(t)$ and the weather forecast $\hat{W}(t+\tau)$, where $\tau$ is the time difference to the prediction target. 
We do not have weather data from the station directly, but can use data from a weather station nearby (ID: 10379). The weather forecast data are obtained every six hours at the closest location publicly accessible, 52.5 latitude and 13 longitude. Typical meteorological data contain, but are not limited to, the wind speed and direction, pressure at different levels, the irradiation, cloud coverage, temperature and humidity. However, in this example, we only use temperature and cloudiness as well as their forecasts as features for our model. The latter is obtained by minimizing
\begin{equation}
\begin{aligned}
\Gamma_1 &=&& \text{NRMSE}\Big(P(t+\tau), \hat{P}\big(P(t), W(t), \hat{W}(t+\tau)\big)\Big) \\
\Gamma_2 &=&& \text{size}(f)
\end{aligned}
\end{equation} 
with $f$ the model under consideration. Our prediction target is $\hat{P}(t +\tau)$ with $\tau= 24$, the one-day-ahead power production. We create our datasets with a sampling of 1h. While additional information from the solar power data remains unused, the prediction variables have to be interpolated. The quality of the forecast depends on quality of the weather measurement and weather forecast. As we use publicly available data, we can only demonstrate the procedure and cannot attain errors as low as those used in commercial products, which will be discussed elsewhere. The features of the the data set are listed in Table~\ref{table:features}
\begin{table*}
\begin{tabular}{|l|l|l|l|l|}
\hline 
Name & Symbol & Source & Sampling & Variable\\
\hline
Solar power & $P(t)$ & direct access & 10 min & $x_1$\\
Total cloud coverage & $\text{tcc}(t)$ & Synop & 1h & $x_4$\\
2 meter temperature & $T(t)$ & Synop & 1h & $x_3$\\
Total cloud coverage prediction & $\text{tcc}_{pred}(t, \tau)$ & ECMWF-TIGGE & 6h & $x_2$\\
2 meter temperature prediction& $T_{pred}(t, \tau)$ & ECMWF-TIGGE & 6h & $x_0$\\
\hline
\end{tabular} 
\centering
\label{table:features}
\caption{Solar power study: description of the data set. We use a set of 5 features drawn from different sources with different sampling.}
\end{table*}
Furthermore, we scale each feature to have its minimum equal zero and maximum equal to one. The models are trained with data from June and July of 2014. Testing is conducted for August 2014.
To obtain first impression (assuming no prior knowledge), we calculate the mutual correlation of the data. The power produced the next day is heavily correlated with the predicted solar irradiation. This is a confirmation that the physics involved is mirrored in the model and that weather prediction is good in average. Quantitative statements on the quality of weather prediction is not easy and can be found in the literature \cite{ecmwf}.

\subsubsection{GP Results}

\begin{figure}
 \includegraphics[width=0.9\linewidth]{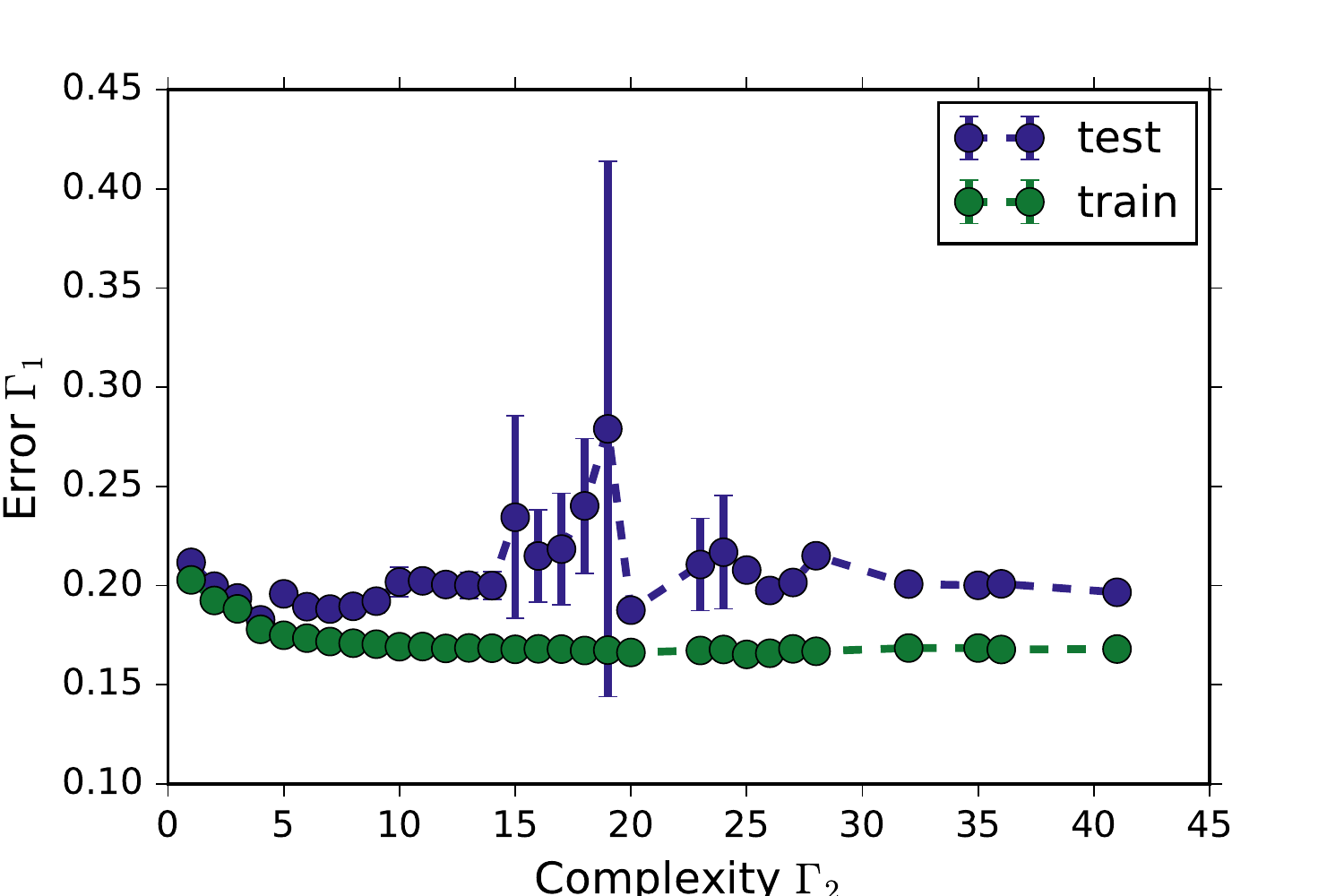}
 \caption{Solar power study, Average Pareto front obtained using GP: with increasing complexity, training and testing data first behave similarly, then testing deviates strongly indicating overfitting or too small testing data set, respectively. The peaks around complexity 20 are due to two reasons: there are only few models on the fronts (1-3), and one of them is an extreme outlier.}
 \centering
 \label{fig:gpPareto}
\end{figure}

Let us consider the results of our forecasting with GP shown in Fig.~\ref{fig:gpPareto}. The Pareto fronts are shown for both the training and testing set. As above, for the coupled oscillators, we have conducted 10 runs with different seeds and display the averaged result. Of course, for the training set (filled diamonds), increasing complexity means decreasing error. We see a strong deviation for very complicated models of the testing data (filled circles). This may be an indication of a small testing sample, or indicate overfitting. The outlier at $\Gamma=18$ is a result of the particular realization of the evolutionary optimization. With a different setting, e.g. more iterations, or multiple runs such outliers are eliminated.
To clarify this question, we show the functions found as solution of our procedure with increasing complexity and one specific seed (42) in Table~\ref{table:solar_results_gp}.

\begin{table*}
\begin{tabularx}{\linewidth}{|l|l|X|}
\hline
Error $\Gamma_1$ & Complexity $\Gamma_2$ & Formula \\
\hline
0.2117 & 1.0 & $x_{1}$\\
\hline
0.1997 & 2.0 & $\sin(x_{1})$\\
\hline
0.1938 & 3.0 & $\sin(\sin(x_{1}))$\\
\hline
0.1827 & 4.0 & $0.662\sqrt(x_{1})$\\
\hline
0.1993 & 5.0 & $\sqrt(x_{0}\sin(x_{1}))$\\
\hline
0.1931 & 6.0 & $\sqrt(\sin(x_{0})\sin(x_{1}))$\\
\hline
0.189 & 7.0 & $\sqrt(x_{0}x_{1}\cos(x_{3}))$\\
\hline
0.1943 & 8.0 & $\sqrt(x_{1})(x_{0} - x_{3} + 0.649)$\\
\hline
0.2348 & 9.0 & $\sqrt(x_{1})\cos(x_{2}x_{3}/x_{0})$\\
\hline
0.192 & 10.0 & $\sqrt(x_{1})(x_{0} - x_{4}(x_{3} - 0.699))$\\
\hline
0.2057 & 12.0 & $\sqrt(x_{1})(x_{0} - x_{2}(x_{2}x_{3} - 0.592))$\\
\hline
0.2684 & 16.0 & $\sqrt(x_{1})(x_{0} + (-x_{2}x_{3} + 0.597)\sin(x_{4} + \sin(x_{1})))$\\
\hline
0.1995 & 18.0 & $-x_{0}\sqrt(x_{1})((\sin(x_{3}) - 0.641)\exp(x_{4})\cos(x_{3}) - 1)$\\
\hline
0.1904 & 25.0 & $x_{0}(\sqrt(x_{1}) + (-x_{3} + 0.715)(\sin(x_{1}) + \sin(x_{2}x_{4}))\cos(x_{1}))$\\
\hline
\end{tabularx}
\caption{Solar power study, method GP: formulae of the Pareto front models for seed 42.}
\label{table:solar_results_gp}
\end{table*}

From Table~\ref{table:solar_results_gp} we see that GP follows a very reasonable strategy: First, it recognizes that the persistence method is a very reasonable thing, with production tomorrow being the same as today ($x_1=P(t)$). Veto a complexity of 5, the identified models only depend on the solar power $x_1$ and describe with increasing accuracy the conditioned average daily profile. The more complex models include the weather data and forecast. The geometric mean of current power and predicted temperature is present. However, due to the low quality weather forecast as well as the seasonal weather difference between training and testing data, there is no net gain in prediction quality.

\begin{figure}
 \subfloat[]{\includegraphics[width=0.9\linewidth]{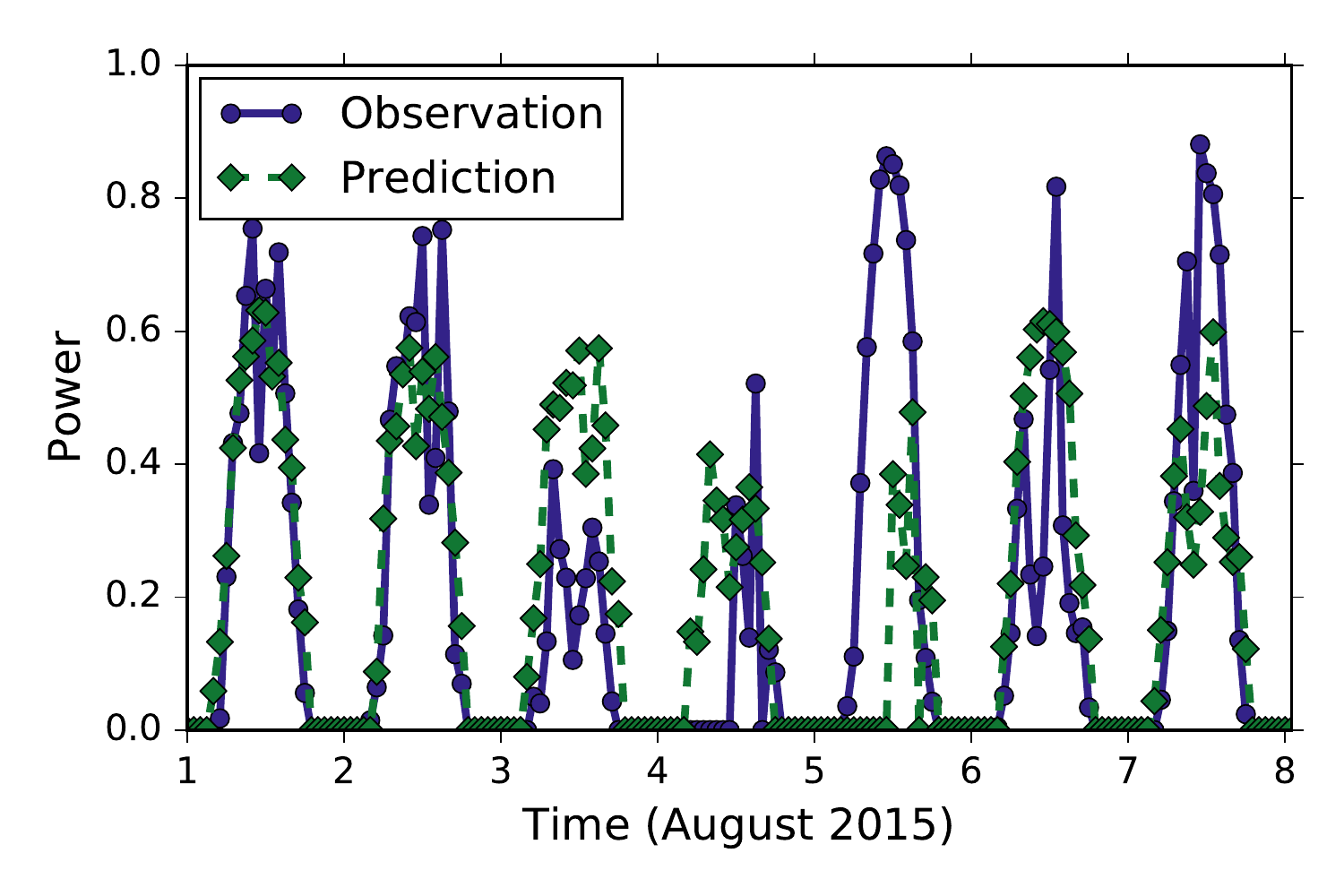}} \\
 \subfloat[]{\includegraphics[width=0.9\linewidth]{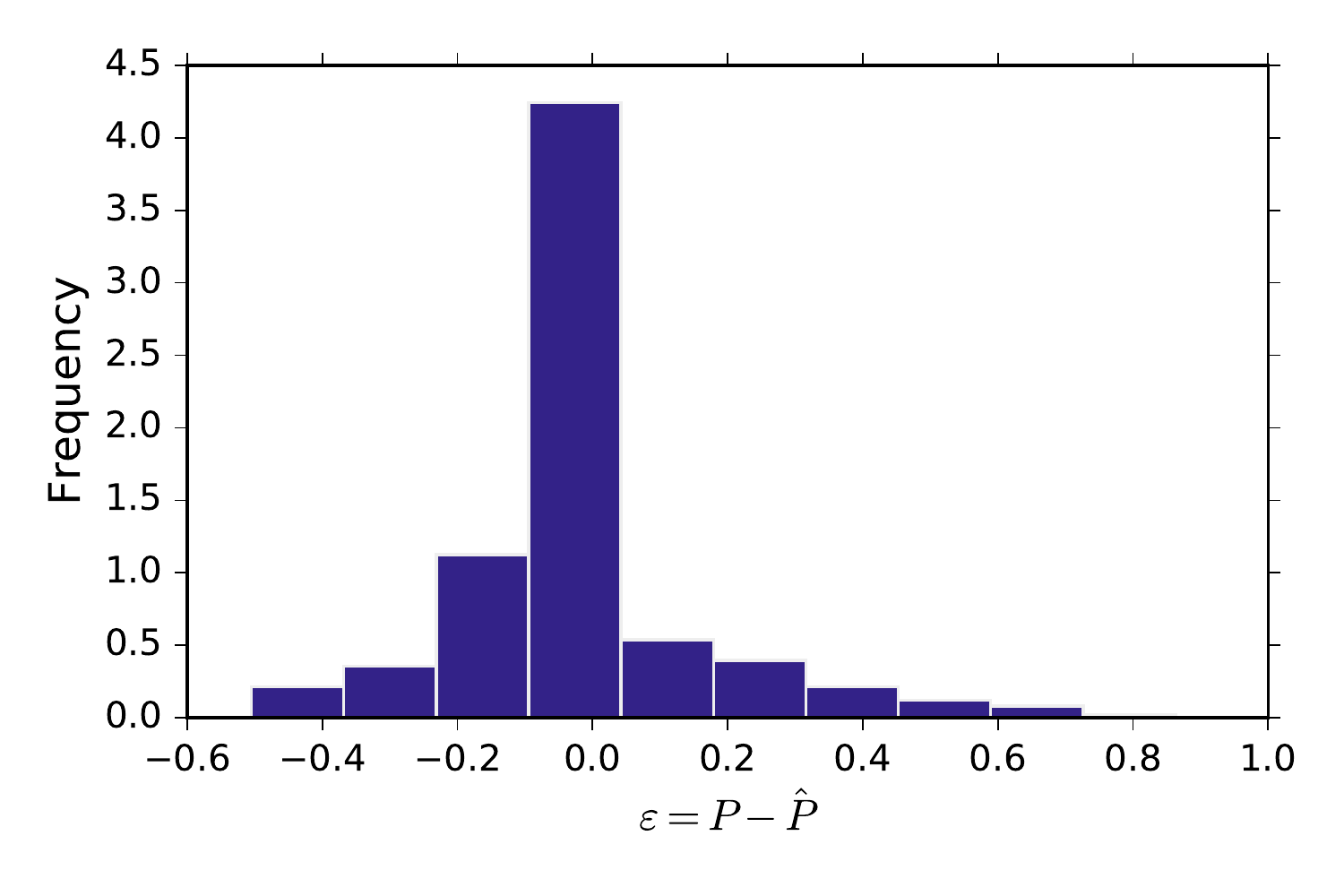}}
 \caption{Solar power study, method GP: a) Real and predicted time series. We display the results of the first week of August 2015. Prediction used model of complexity 4 which had lowest error on the test set. b) Histogram of the residuals $\varepsilon = P - \hat{P}$. The distribution is asymmetric around zero. The model tends to underpredict.}
 \centering
 \label{fig:gpresult}
 \end{figure}

Without any further analysis, the model with the lowest testing error is chosen. In Fig.~\ref{fig:gpresult} (a) we confront the real time series with the prediction from GP for the model of complexity 4. 
One clearly finds the already mentioned conditioned average profile. This predicts the production onset a bit too early. The error distribution is shown in Fig.~\ref{fig:gpresult} (b), where we recognize an asymmetric error distribution with more probable under- than overprediction. 

\subsubsection{FFX Results}

\begin{figure}
 \includegraphics[width=0.9\linewidth]{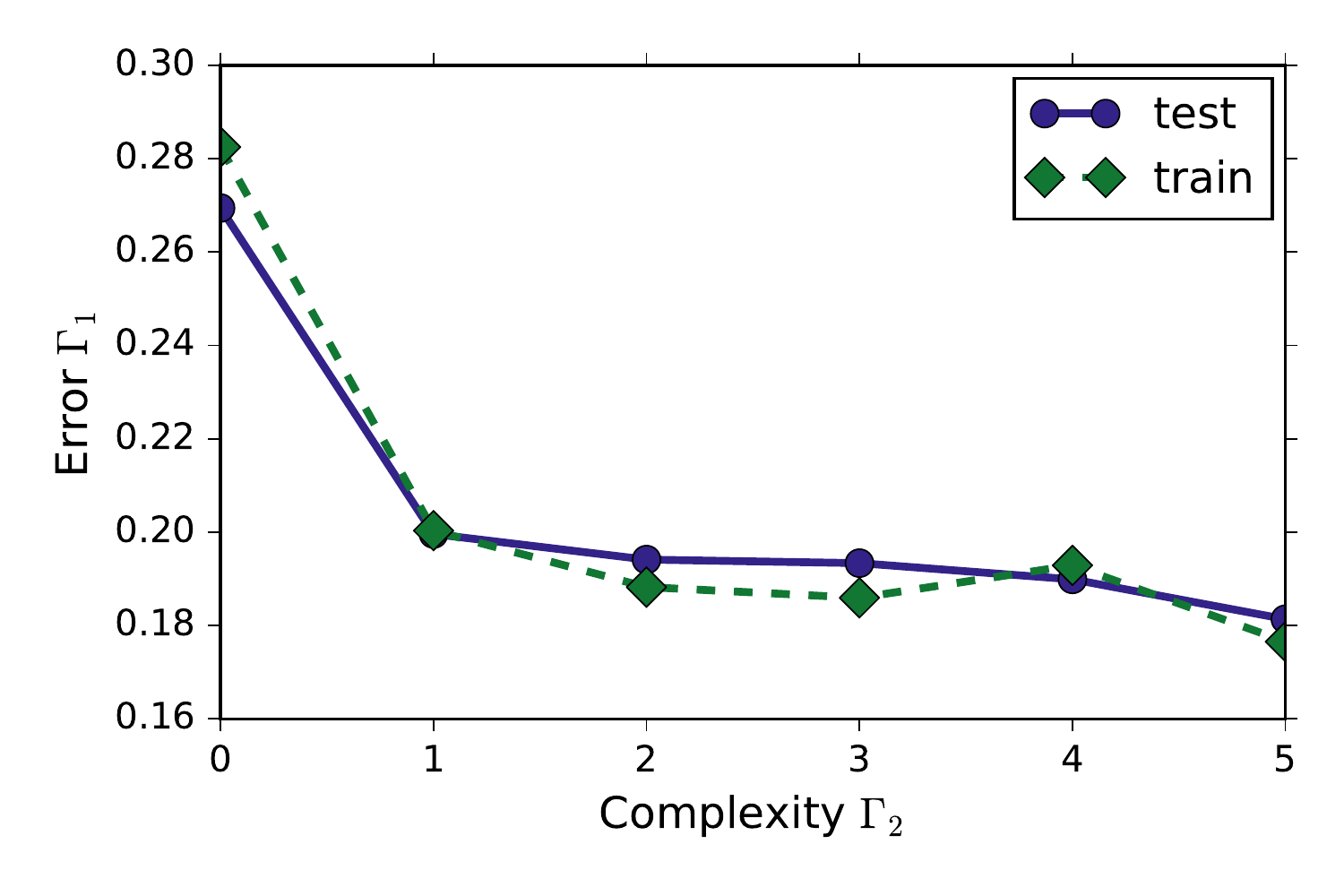}
 \caption{Solar power study, Pareto front obtained using FFX. The results for FFX are as accurate as the ones obtained with GP. Test and training set are, however, nicely aligned. This demonstrates not only consistency of the models, but less variability of the models found.} 
 \centering
 \label{fig:ffx_pareto_solar}
\end{figure}

\begin{figure}
 \subfloat[]{\includegraphics[width=0.9\linewidth]{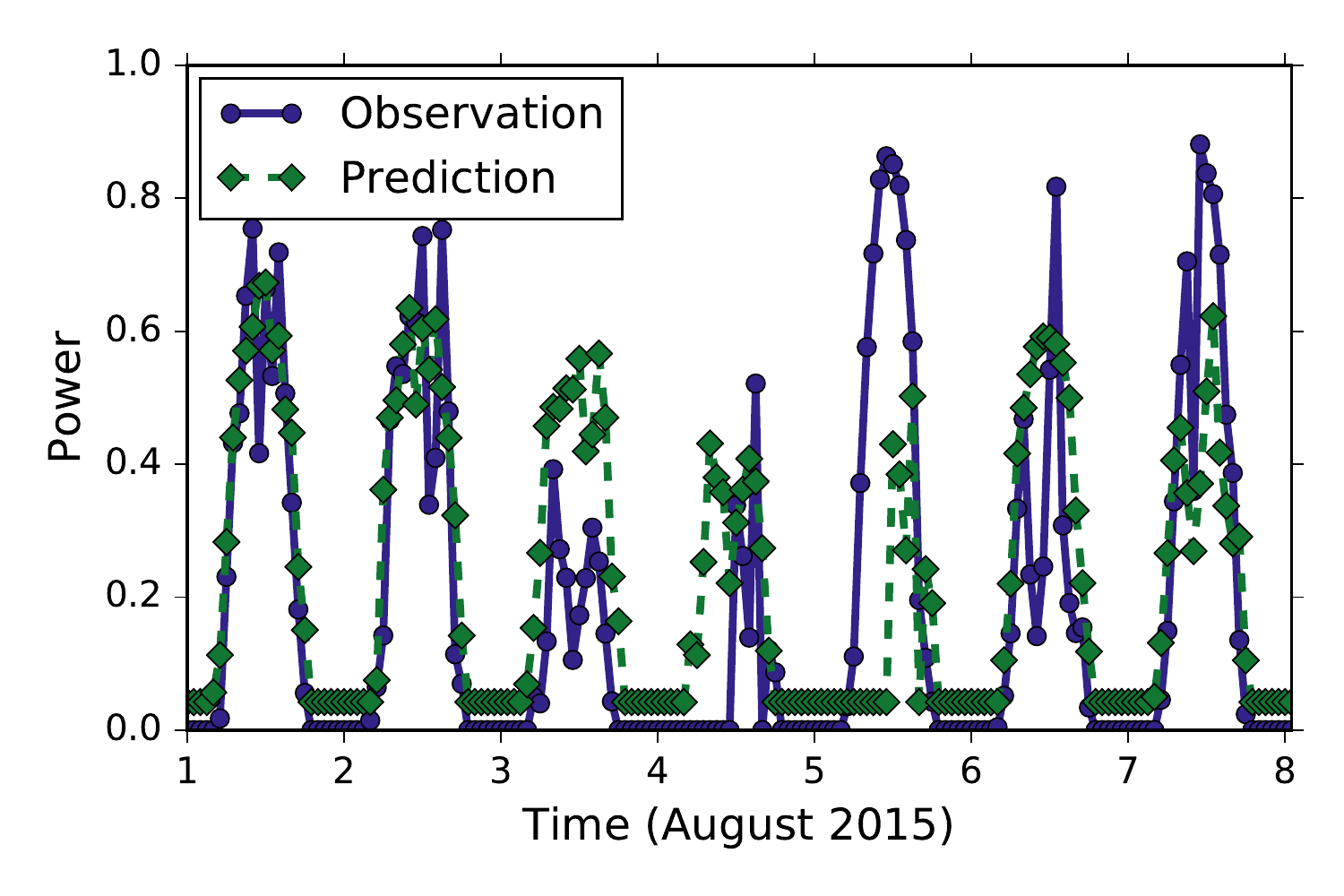}} \\
 \subfloat[]{\includegraphics[width=0.9\linewidth]{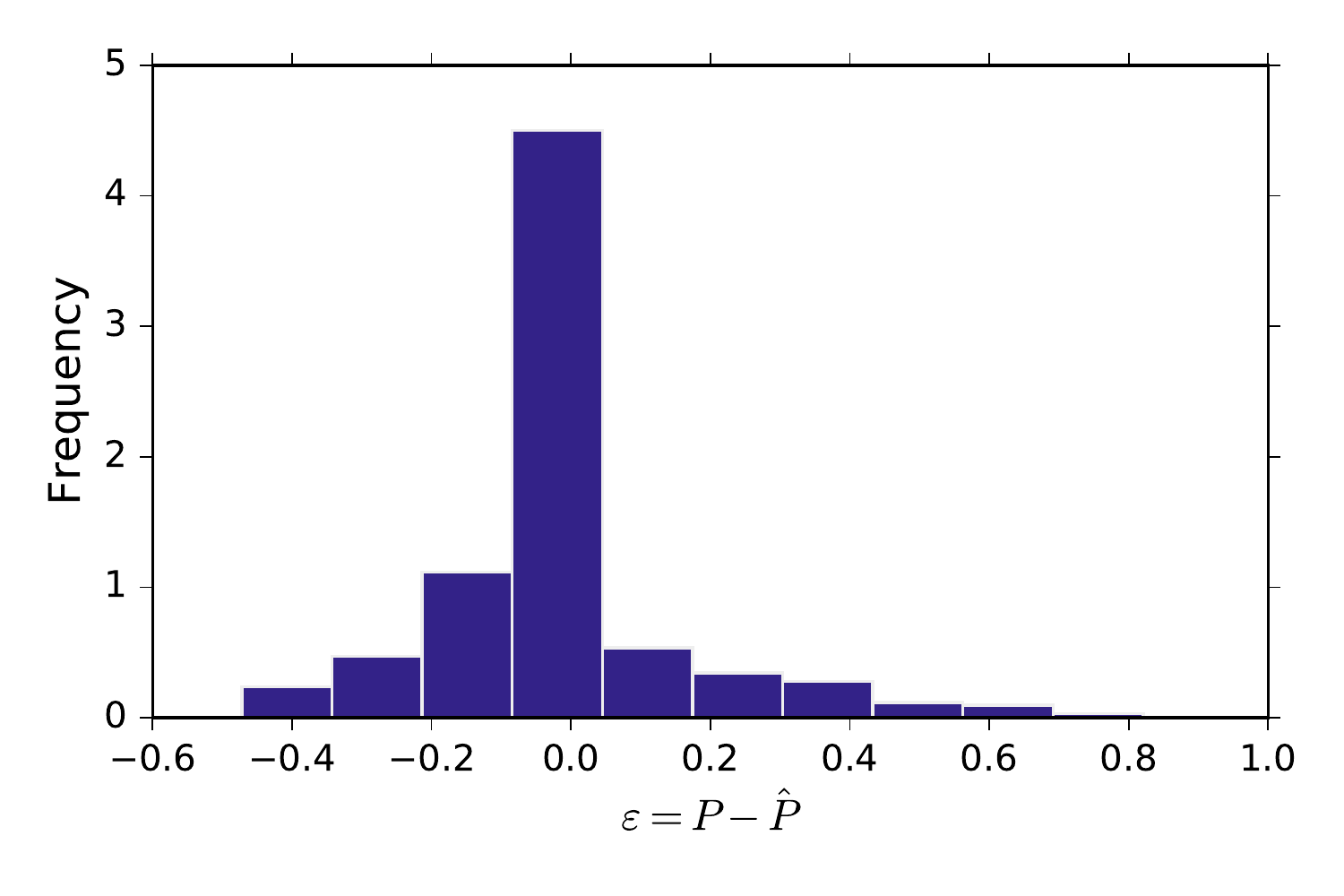}}
 \caption{Solar power study, method FFX: a) Timeseries of the predicted ($\hat{P}$), and observed ($P$) data. We display the results of the first week of August 2015. Similar to the GP prediction extrema are not particularly well predicted. For the linear model, even the zero values are not well hit. The reason for this is the regression to mean values and the inability of powers to stay at zero for a sufficient time. b) Histogram of the residuals $\varepsilon = P - \hat{P}$. Despite different formulas, the histogram of the residuals is asymmetric around zero with a trend to underpredict as well.}
 \centering
 \label{fig:ffxScatterAndTS}
\end{figure}

The results of the FFX method are shown in Fig.~\ref{fig:ffx_pareto_solar}-\ref{fig:ffxScatterAndTS} and the models in Table~\ref{table:solar_results_ffx}.
As shown, the FFX method is less capable of predicting longer inactive periods, such as at night, where no solar power is produced. This is clearly visible in Fig.~\ref{fig:ffxScatterAndTS}. 

Analyzing the equations of Table~\ref{table:solar_results_ffx}, we notice that the best FFX function is a quadratic form with maxima to limit the signal above zero. This amounts to recover the mean shape of the signal as a quadratic function. Unfortunately this seems almost trivial since one could obtain this mean shape by purely geometrical considerations with a factor for the cloud coverage.

\begin{table*}
\begin{tabularx}{\linewidth}{|l|l|X|}
\hline
Error $\Gamma_1$ & Complexity $\Gamma_2$ & Formula \\
\hline
0.269419547644 & 0 & $0.221$\\
\hline
0.199597415208 & 1 & $0.108 + 0.511x_{1}$\\
\hline
0.194122751788 & 2 & $0.0223 + 0.606x_{1} + 0.139x_{0}$\\
\hline
0.193361252172 & 3 & $0.0470 + 0.436x_{1} + 0.328x_{0}  x_{1} + 0.138x_{4}  x_{0}$\\
\hline
0.189889358594 & 4 & $0.459 - 0.458\max(0,0.200-x_{1}) - 0.339\max(0,0.333-x_{1}) - 0.134\max(0,0.733-x_{1}) - 0.0828\max(0,0.867-x_{1})$\\
\hline
0.18135731346 & 5 & $0.301 - 1.25\max(0,0.333-x_{1})  \max(0,0.200-x_{1}) - 0.810\max(0,0.467-x_{1})  \max(0,0.200-x_{1}) + 0.457x_{0}  x_{1} - 0.252\max(0,0.333-x_{1}) - 0.0794\max(0,0.200-x_{1})$\\
\hline
\end{tabularx}
\caption{Solar power study, method FFX: formulae of the Pareto front models.}
\label{table:solar_results_ffx}
\end{table*}

Summarizing the results for the solar power curves, both methods are able to reproduce the true curve to approximately 20\% which is reasonable for a nonoptimized method. The detection of changes when clear sky switches to partially or fully clouded one is not entirely satisfactory and one needs to investigate the improvement of weather predictions for a single location. As said in the introduction, a perfect weather prediction with high resolution would render this work useless for power production forecast  (although not for other questions).

Nevertheless, we note that the results in the form of analytic models are highly valuable, because interpretations and further mathematical analysis are possible.

\section{Conclusion}
\label{sec:conclusion}

We have demonstrated the use of symbolic regression combined with complexity analysis of the resulting models for the future prediction of dynamical systems. More precisely, we identify a system of equations yielding optimal forecasts in terms of a minimized normalized root mean squared error of the difference between model forecast and observation of the system state. We did not investigate theoretical aspects such as the underlying state space, nor what implications of the functions on the model. These will be subject of future investigations. Such work is to be carried out carefully to find the limitations of the approach, in particular of genetic programming, which is rather uncontrolled in the way the search space is explored. On the other hand, the methods stand in line with a large collection of methods from regression and classification and one can use much of this previous knowledge. In our opinion, the multiobjective analysis is crucial to identify models to a degree such that they can be used in practice. Probably, this approach will prove very helpful if used in combination with scale analysis, e.g. by prefiltering the data on a selected spatio-temporal scale and then identify equations for this level. 

We have tried to show the possible power by three examples of increasing complexity: a trivial one - the harmonic oscillator with an almost perfect predictive power, a collection of excitable oscillators where we demonstrated that the methods can perform a kind of multi-scale analysis based on the data.
Thirdly, examine the one-day-ahead forecasting of solar power production we have shown that even for messy data we can improve the classical methods by a few percent (in NRMSE).
For theoretical considerations, this might be negligible, for real world applications, a few percent might translate into a considerable advantage, since the usage of rare resources can be optimized.

A question for further research is how we can use simplification during the GP iteration to alter the complexity. It may be even a viable choice to control the complexity growth over time, the so-called bloat, in single objective genetic programming - a topic of ongoing interest \citep{gardner2014controlling}. Additionally, we introduced an intermediate step to only allow for one of many identical solutions for further evolution. One could consider to expand the idea of identical expression trees to include symmetries. 

We conclude that symbolic regression is very useful for the prediction of dynamical systems, based on observations only. Our future research will focus on the use of equations couple the systems to other macroscopic ones (e.g. finance, in the case of wind power), and on the analysis of system stability and other fundamental properties using the found equations, which is scientifically a very crucial point.

\section*{Acknowledgements}

We acknowledge discussion with C. Cornaro and M. Pierro on solar power forecast, helpful words on machine learning by M. Segond. We thank the unisolar initiative for their provision of the data. M. Quade and M. Abel acknowledge support by the German ministry of economy, ZIM project "green energy forecast", grant ID KF2768302ED4.

\end{document}